\documentclass[twocolumn,twocolappendix]{aastex631}
\usepackage{amsmath}
\usepackage{tcolorbox}

\def\Fig{\mbox{Figure~}}

\def\Tab{\mbox{Table~}}

\def\Sec{\mbox{Section~}}
\def\Secs{\mbox{Sections~}}
\def\Eq{\mbox{Equation~}}
\def\Eqs{\mbox{Equations~}}

\begin{document}

\title{Illuminating the Dark Side of Cosmic Star Formation III:\\
Building the largest homogeneous sample of Radio-Selected Dusty Star-Forming Galaxies in COSMOS with PhoEBO}

\correspondingauthor{Fabrizio Gentile}
\email{fabrizio.gentile3@unibo.it}

\author[0000-0002-8008-9871]{Fabrizio Gentile}
\affiliation{University of Bologna, Department of Physics and Astronomy (DIFA), Via Gobetti 93/2, I-40129, Bologna, Italy}
\affiliation{INAF -- Osservatorio di Astrofisica e Scienza dello Spazio, via Gobetti 93/3 - 40129, Bologna - Italy}

\author[0000-0003-4352-2063]{Margherita Talia}
\affiliation{University of Bologna, Department of Physics and Astronomy (DIFA), Via Gobetti 93/2, I-40129, Bologna, Italy}
\affiliation{INAF -- Osservatorio di Astrofisica e Scienza dello Spazio, via Gobetti 93/3 - 40129, Bologna - Italy}

\author[0000-0002-6444-8547]{Meriem Behiri}
\affiliation{SISSA, Via Bonomea 265, I-34136 Trieste, Italy}

\author[0000-0002-2318-301X]{Giovanni Zamorani}
\affiliation{INAF -- Osservatorio di Astrofisica e Scienza dello Spazio, via Gobetti 93/3 - 40129, Bologna - Italy}

\author[0000-0003-3419-538X]{Luigi Barchiesi}
\affiliation{University of Bologna, Department of Physics and Astronomy (DIFA), Via Gobetti 93/2, I-40129, Bologna, Italy}
\affiliation{INAF -- Osservatorio di Astrofisica e Scienza dello Spazio, via Gobetti 93/3 - 40129, Bologna - Italy}

\author[0000-0002-8853-9611]{Cristian Vignali}
\affiliation{University of Bologna, Department of Physics and Astronomy (DIFA), Via Gobetti 93/2, I-40129, Bologna, Italy}
\affiliation{INAF -- Osservatorio di Astrofisica e Scienza dello Spazio, via Gobetti 93/3 - 40129, Bologna - Italy}

\author[0000-0002-7412-647X]{Francesca Pozzi}
\affiliation{University of Bologna, Department of Physics and Astronomy (DIFA), Via Gobetti 93/2, I-40129, Bologna, Italy}
\affiliation{INAF -- Osservatorio di Astrofisica e Scienza dello Spazio, via Gobetti 93/3 - 40129, Bologna - Italy}

\author[0000-0002-3915-2015]{Matthieu Bethermin}
\affiliation{Universit\'e de Strasbourg, CNRS, Observatoire astronomique de Strasbourg, UMR 7550, 67000 Strasbourg, France}
\affiliation{Aix Marseille Univ, CNRS, CNES, LAM, Marseille, France}

\author[0000-0002-0200-2857]{Andrea Enia}
\affiliation{University of Bologna, Department of Physics and Astronomy (DIFA), Via Gobetti 93/2, I-40129, Bologna, Italy}
\affiliation{INAF -- Osservatorio di Astrofisica e Scienza dello Spazio, via Gobetti 93/3 - 40129, Bologna - Italy}

\author[0000-0002-9382-9832]{Andreas L. Faisst}
\affiliation{Caltech/IPAC, MS314-6, 1200 E. California Blvd. Pasadena, CA, 91125, USA}

\author[0000-0002-1847-4496]{Marika Giulietti}
\affiliation{SISSA, Via Bonomea 265, I-34136 Trieste, Italy}
\affiliation{INAF -- Osservatorio di Astrofisica e Scienza dello Spazio, via Gobetti 93/3 - 40129, Bologna - Italy}

\author[0000-0002-5836-4056]{Carlotta Gruppioni}
\affiliation{INAF -- Osservatorio di Astrofisica e Scienza dello Spazio, via Gobetti 93/3 - 40129, Bologna - Italy}

\author[0000-0002-4882-1735]{Andrea Lapi}
\affiliation{SISSA, Via Bonomea 265, I-34136 Trieste, Italy}
\affiliation{IFPU - Institute for fundamental physics of the Universe, Via Beirut 2, 34014 Trieste, Italy}
\affiliation{INFN-Sezione di Trieste, via Valerio 2, 34127 Trieste, Italy}
\affiliation{INAF/IRA, Istituto di Radioastronomia, Via Piero Gobetti 101, 40129 Bologna, Italy}

\author[0000-0002-0375-8330]{Marcella Massardi}
\affiliation{INAF/IRA, Istituto di Radioastronomia, Via Piero Gobetti 101, 40129 Bologna, Italy}
\affiliation{INAF, Istituto di Radioastronomia - Italian ARC, Via Piero Gobetti 101, I-40129 Bologna, Italy}
\affiliation{SISSA, Via Bonomea 265, I-34136 Trieste, Italy}

\author[0000-0002-5836-4056]{Vernesa Smol{\v{c}}i{\'c}}
\affiliation{Department of Physics, University of Zagreb, Bijenicka cesta 32, 10002 Zagreb, Croatia}

\author[0000-0002-6748-0577]{Mattia Vaccari}
\affiliation{Inter-University Institute for Data Intensive Astronomy, Department of Astronomy, University of Cape Town, 7701 Rondebosch, Cape Town, South Africa}
\affiliation{Inter-University Institute for Data Intensive Astronomy, Department of Physics and Astronomy, University of the Western Cape, 7535 Bellville, Cape Town, South Africa}
\affiliation{INAF/IRA, Istituto di Radioastronomia, Via Piero Gobetti 101, 40129 Bologna, Italy}

\author[0000-0002-4409-5633]{Andrea Cimatti}
\affiliation{University of Bologna, Department of Physics and Astronomy (DIFA), Via Gobetti 93/2, I-40129, Bologna, Italy}
\affiliation{INAF -- Osservatorio di Astrofisica e Scienza dello Spazio, via Gobetti 93/3 - 40129, Bologna - Italy}

\begin{abstract}
\noindent In the last decades, an increasing scientific interest has been growing in the elusive population of “dark" (i.e. lacking an optical/NIR counterpart) Dusty Star-Forming Galaxies (DSFGs). Although extremely promising for their likely contribution to the cosmic Star Formation Rate Density and for their possible role in the evolution of the first massive and passive galaxies around $z\sim3$, the difficulty in selecting statistically significant samples of dark DSFGs is limiting their scientific potentialities.
This work presents the first panchromatic study of a sample of { 263} Radio-Selected NIRdark galaxies discovered in the COSMOS field following the procedure by \citet{Talia_21}. These sources are selected as radio-bright galaxies ($S_{\rm 3GHz}>12.65\mu$Jy) with no counterpart in the NIR-selected COSMOS2020 catalog ($Ks \gtrsim$ 25.5 mag). For these sources, we build a new photometric catalog including accurate photometry from the optical to the radio obtained with a new deblending pipeline (\textsc{PhoEBO}: Photometry Extractor for Blended Objects). We employ this catalog to estimate the photo-\textit{z}s and the physical properties of the galaxies through an SED-fitting procedure performed with two different codes (\textsc{Magphys} and \textsc{Cigale}). Finally, we estimate the AGN contamination in our sample by performing a series of complementary tests. The high values of the median extinction ($A_v\sim4$) and star formation rate (${\rm SFR}\sim500$ M$_\odot$/yr) confirm the likely DSFG nature of the RS-NIRdark galaxies. The median photo-\textit{z} ($z\sim3$) and the presence of a significant tail of high-\textit{z} candidates ($z>4.5$) suggest that these sources are important contributors to the cosmic SFRD and the evolutionary path of galaxies at high redshifts.
\end{abstract}


\section{Introduction}
\label{sec:intro}

\noindent Our picture of the Universe is essentially the offspring of the instruments we use to observe it. Until a few years ago, the Hubble Space Telescope (HST) was the most powerful facility available to study the high-\textit{z} Universe. The data from this facility have put several critical constraints on important quantities as a function of cosmic time (see, e.g., the review by \citealt{Madau_14} and references therein). One of the main results concerned the reconstruction of the cosmic Star Formation Rate Density (SFRD; i.e. the amount of stellar mass formed in the Universe per each year and Mpc$^3$) back to the epoch of reionization ($z\sim8$). The vast majority of these studies agreed in describing a SFRD constantly growing $\propto (1+z)^{1.7}$ from $z\sim0$ up to $z\sim2.5$ (the so-called \textit{cosmic noon}) and - then - decreasing $\propto (1+z)^{-2.9}$ up to $z\sim8$ \citep{Madau_14}. Given the spectral coverage of HST, however, our census of galaxies in the first few billion years of cosmic history was forcedly limited to optically-bright galaxies. Given the redshift ranges explored, these sources were mainly rest-frame UV{ /optically}-bright galaxies. 

Even in that time, however, an increasing number of evidences suggested that our picture of galaxy evolution based on these UV-bright sources was quite far from complete. Above all, the discovery of a significant population of massive ($M_\star>10^{11} M_\odot$) and passive (sSFR$<10^{-11} {\rm yr}^{-1}$) galaxies already in place at $z\sim3$ (i.e. when the Universe was just { 2} Gyr old; see some examples in \citealt{Straatman_14} and \citealt{Schreiber_18}) was – probably – the most outstanding. The number density of these sources ($n\sim2\times 10^{-5}$ Mpc$^{-3}$) resulted in being more than two orders of magnitude higher than that computed for UV-bright galaxies at $z>4.5$ \citep[see e.g. the discussion in ][]{Straatman_14,Schreiber_18,Valentino_20}. This suggested that, when not accounting for UV-dark star forming galaxies,  we are missing a significant fraction of the progenitors of high-\textit{z} passive galaxies.

Observations carried out with facilities able to explore other regions of the electromagnetic spectrum such as the sub-millimeter array camera SCUBA on the James Clerk Maxwell Telescope \citep[e.g.][]{Smail_97,Hughes_98} or the Herschel Space Observatory \citep[e.g.][]{Gruppioni_13,Burgarella_13}  confirmed that selections based only on the optical regime actually miss a significant fraction of high-\textit{z} galaxies (see e.g. \citealt{Blain_02,Weiss_13}). Among the main missing population of galaxies, the so-called “Dusty Star-Forming Galaxies" (DSFGs; see e.g. \citealt{Lagache_05,casey_14} for reviews) are worth a deeper discussion. These sources are characterized by high values of the Star Formation Rate and high stellar extinction due to the presence of a significant amount of dust. Hence, these sources are extremely faint or even undetected in the optical/NIR regime observed by HST. Despite their elusiveness, several studies targeting these sources \citep[e.g.][]{Wang_19,Gruppioni_20,Talia_21,Zavala_21,Enia_22,VanDerVlugt_22} are unanimous in finding that the contribution of these galaxies to the cosmic SFRD in the high-\textit{z} regime is definitely significant (up to 40\% the contribution of optically bright galaxies at $z>4.5$; see e.g. \citealt{Talia_21,Enia_22,Behiri_23}). The inclusion of these new sources in the cosmic census of the high-\textit{z} galaxies could even change the behavior of the SFRD at $z>3$ (favoring a flatter behavior before the cosmic noon; see e.g. \citealt{Gruppioni_20}) and solve the puzzle of the missing progenitors of massive galaxies \citep[e.g.][]{Toft_14}.

It is crucial to underline how all the aforementioned studies on DSFGs are generally based on limited samples of galaxies. The main reason for this is the extreme faintness of these sources in the optical/NIR regime, making their identification feasible only at longer wavelengths. For this reason, these galaxies are generally selected in the IR or (sub)mm regimes \citep[e.g.][]{Smail_21,Dudzevicute_21,Manning_22,Giulietti_22}, taking advantage of the bright thermal emission of the dust and the intensely negative \textit{k}-correction that makes these sources bright even at high redshifts.

IR/(sub)mm selections, however, are affected by - at least - two significant issues. The first one is the low sensitivity of single-dish (sub)mm telescopes. This, combined with their large beam sizes, makes it hard to identify the multi-wavelength counterparts of each galaxy. In principle, this problem could be overcome by employing state-of-the-art facilities such as the \textit{Atacama Large (sub)Millimeter Array} (ALMA) with higher sensitivities and smaller beams. However, the deep surveys conducted with these telescopes \citep[e.g.][]{Franco_18,Casey_21} are currently mapping small volumes, hence are affected by low statistics and are prone to cosmic variance. The second issue concerning IR/(sub)mm selections resides in the still-debated properties of the dust in the high-\textit{z} regime. For instance, a correlation between the dust temperature and the redshift (e.g. the one suggested by \citealt{Bethermin_15}, \citealt{Faisst_17} and \citealt{Schreiber_18b}) could produce a significant selection bias.

A solution to both the aforementioned issues can reside in the radio selection of DSFGs \citep[e.g.][]{Talia_21,Enia_22,Behiri_23}. In “normal" (i.e. non-active) galaxies, radio photons can be generated by free-free emission in HII regions and synchrotron emission from relativistic electrons accelerated in supernovae remnants. In addition, they are not affected by the presence of dust. For these reasons, once excluded the nuclear origin of these photons, they could represent an excellent unbiased tracer of star formation. Therefore, as noted by \citet{Talia_21}, the selection of radio-bright galaxies with a faint or undetected counterpart in the optical/NIR regime could provide a significant sample of likely DSFGs.

This paper expands the studies by \citet{Talia_21} and \citet{Behiri_23}, which aimed to assemble the widest homogeneously-selected sample of candidate DSFGs available in the current literature. We take advantage of the multi-wavelength coverage in the Cosmic Evolution Survey (COSMOS; \citealt{Scoville_07}) field and the high sensitivity of the VLA-3GHz COSMOS Large Project \citep{Smolcic_17} to collect a sample of 323 Radio-Selected NIRdark Galaxies (RS-NIRdark galaxies in the following). In this paper, we present the photometric catalog of these sources with new accurate photometry extracted from the optical to the radio regimes. We also employ a SED-fitting procedure to estimate photo-\textit{z}s and physical properties of our galaxies.

The paper is structured as follows. In \Sec\ref{sec:selection}, we introduce the sample selection. In \Sec\ref{sec:data}, we describe the maps employed to extract the photometry in the optical-to-MIR regime and the additional photometry retrieved at longer wavelengths. In \Sec\ref{sec:photometry_extraction}, we present the new \textsc{PhoEBO} pipeline (Photometry Extractor for Blended Objects) used to extract the photometry and its validation on simulated data. \Sec\ref{sec:properties} is focused on the SED-fitting procedure employed to assess the photo-\textit{z}s and the physical properties of the galaxies in our sample. Furthermore, in \Sec\ref{sec:AGN} we analyze in detail the possible AGN contamination in our sample. In \Sec\ref{sec:discussion}, we discuss what the results presented in this paper tell us about the nature of the RS-NIRdark galaxies and compare them with previous analogous studies in the current literature. Finally, we draw our conclusions in \Sec\ref{sec:summary}. Throughout this paper, we use AB magnitudes \citep{Oke_83}, employ a \citet{Chabrier_03} Initial Mass Function (IMF) and assume a concordance cosmology $\Lambda$CDM with $H_0=70$ km s$^{-1}$ Mpc$^{-1}$ and ($\Omega_{\rm tot}, \Omega_{\Lambda}, \Omega_m$)=(1,0.7,0.3).

\section{Sample selection}
\label{sec:selection}

\noindent To assemble a large sample of candidate DSFGs with complete photometry from the optical to the radio, we focus on the galaxies in the COSMOS field. We collect our sources by performing a selection analogous to that employed by \citet{Talia_21} and summarized here:

\begin{enumerate}
\item We start from the VLA-COSMOS 3GHz Large Project catalog \citep{Smolcic_17}. We select 8850 radio sources with an S/N$>5.5$ (i.e brighter than 12.65 $\mu$Jy beam$^{-1}$) over the full 2 deg$^2$ coverage of the survey. This cut allows us to assemble a sample of radio-bright galaxies, limiting the likely contamination of fake sources to 0.4\% \citep{Smolcic_17}.
\item To limit the expected presence of galaxies hosting AGN (common in radio-selected catalogs; see e.g. \citealt{Bonzini_13} and \citealt{Novak_18}), we remove from the sample all the sources flagged as “multi-component” in the initial catalog. We underline that this flag is the result of a visual inspection of the sources in the full catalog by \citet{Smolcic_17}. Therefore - at this stage - we cannot exclude that some multi-component radio sources are still present in our sample (see the discussion in \Sec\ref{sec:AGN}).
\item To take advantage of the multi-wavelength coverage of the COSMOS field, we exclude from the sample all the radio sources lying outside the UltraVISTA survey footprint \citep{Laigle_16}. This further limits the sample to 5982 galaxies and the effective area mapped by this study to 1.38 deg$^2$.
\item { Finally, we cross-match the resulting sample with the two versions of the COSMOS2020 catalog (the “\textsc{Classic}" and “\textsc{Farmer}"; see \citealt{Weaver_22}), removing all the sources with a match within 0.7''. The final sample is composed of 323 galaxies}.
\end{enumerate}

The last step aims to include in the catalog only sources without a significant counterpart in the optical/NIR bands. However, it is crucial to underline how the source detection in the COSMOS2020 catalog is performed on the $\chi^2$-image produced through the \textsc{SWarp} software \citep{Bertin_SWarp} by combining the maps in the $i$ and $z$ band from the Subaru telescope and the $Y,J,H$ and $Ks$ bands from the VISTA telescope \citep[see][]{Weaver_22}. The flux in the generic pixel I of this image is computed as a weighted average of the fluxes $f_i$ in the different photometric bands, with the weights $w_i$ provided by the uncertainty maps:
\begin{equation}
\label{eq:chi}
    I=\sqrt{\frac{\sum_{i=1}^N w_i f_i^2}{N}}
\end{equation}
Given this definition, we cannot exclude that some of the sources in our catalog could have a significant counterpart in a few individual NIR bands. { Moreover, through a visual inspection of the $\chi^2$-map employed by \citet{Weaver_22} for detecting the sources in the COSMOS2020, we notice that 60 galaxies of our sample have a detection in that image with a S/N higher than $1.5\sigma$ (i.e. the threshold used in \citealt{Weaver_22}). Some of these sources ($\sim 10\%$) are inside regions masked in the COSMOS2020 that were employed in the COSMOS2015, while most of the lasting sources are close to a bright companion, therefore it is likely that they were not properly deblended in the COSMOS2020. Since these galaxies are not included in the COSMOS2020, they should be part of our sample of RS-NIRdark galaxies. However, due to the different photometry, we expect their properties to differ from the rest of the sample. Therefore, we do not consider these sources in all the statistical analyses performed in this paper aimed to characterize our population of galaxies: for all these studies, we will focus on the lasting 263 galaxies.  { Finally, we underline that the S/N cut applied to the radio catalog excludes from the sample all the faintest radio-sources. Although this step could potentially cause the lack of significantly high-\textit{z} galaxies (see e.g. \citealt{Casey_19}), it is necessary to exclude a high contamination by fake sources (this rate would increase up to 25\% just including the sources with $5<{\rm S/N}<5.5$, see \citealt{Smolcic_17}).}

A significant difference between this study and the previous papers of this series (\citealt{Talia_21} and \citealt{Behiri_23}) resides in the last step of the selection: while the previous works selected a sample of 476 RS-NIRdark galaxies without a counterpart in the COSMOS2015 catalog \citep{Laigle_16}, we employ the updated (and deeper) COSMOS2020 catalog \citep{Weaver_22}. This step allows us to exclude from the sample 153 galaxies un-detected in the COSMOS2015 but revealed by COSMOS2020. { More in detail, 16 sources were only included in the “\textsc{Farmer}" catalog, 21 only in the “\textsc{Classic}", and 116 in both the catalogs.} Finally, improving the works by \citet{Talia_21} and \citet{Behiri_23}, we do not employ any additional selection aimed to avoid sources with a bright contaminant in the vicinity, but we analyze the entire sample, performing a more accurate photometry extraction in presence of contamination or source blending. A more comprehensive comparison with \citet{Talia_21} and \citet{Behiri_23} can be found in Appendix \ref{sec:comparison_t21}.

\section{Data}
\label{sec:data}

\subsection{Analyzed maps}
\label{sec:maps}

\begin{deluxetable*}{ccccccc}
\label{tab:maps}
\tablecaption{Main properties of the maps analyzed in \Sec\ref{sec:data}. { Part of the data are reproduced from \citet{Weaver_22} with author's permission.}}
\tablewidth{0pt}
\tablehead{
\colhead{Instrument} & \colhead{Band} & \colhead{$\lambda^{(a)}$} & \colhead{$\Delta\lambda^{(b)}$ ($\r{A}$)} & \colhead{Depth$^{(c)}$} & \colhead{Corr.} & \colhead{  PSF FWHM$^{(e)}$} \\
\colhead{/Telescope} &  & \colhead{($\r{A}$)} & \colhead{($\r{A}$)} & \colhead{(2'')} & \colhead{Fact.$^{(d)}$} & \colhead{(")}
}
\startdata
& $g$ & 4847 & 1383 & 28.1 & 1.4 & 0.79\\
Hyper-Suprime  & $r$ & 6219 & 1547 & 27.8 & 1.4 & 0.75\\
Cam/Subaru& $i$ & 7699 & 1471 & 27.6 & 1.5 & 0.61\\
& $z$ & 8894 & 766 & 27.2 & 1.4 & 0.68\\
& $y$ & 9761 & 786 & 26.5 & 1.4 & 0.68\\
\hline
 & $Y$ & 10216 & 963 & 25.3/26.6$^{(f)}$ & 2.7/2.8$^{(f)}$ & 0.82 \\
VIRCAM & $J$ & 12525 & 1718 & 25.2/26.4$^{(f)}$ & 2.5/2.7$^{(f)}$ & 0.79 \\
/VISTA & $H$ & 16466 & 2905  & 24.9/26.1$^{(f)}$ & 2.4/2.6$^{(f)}$ & 0.76\\
& $Ks$ & 21557 & 3074 & 25.3/25.7$^{(f)}$ & 2.4/2.4$^{(f)}$ & 0.75\\
\hline
 & ch1 & 35686 & 7443 & 26.4 & - & - \\
IRAC & ch2 & 45067 & { 10119} & 26.3 & - & - \\
/Spitzer & ch3 & 57788 & 14082 & 23.2 & - & -\\
 & ch4 & 79958 & 28796 & 23.1 & - & -\\
\enddata 
\textbf{Notes:} \\
$^{(a)}$ Median $\lambda$ of transmission curve \\
$^{(b)}$ FWHM of the transmission curve \\
$^{(c)}$ $3\sigma$ depths as reported by \citet{Weaver_22} \\
$^{(d)}$ Multiplicative correction for the photometric uncertainties \citep{Weaver_22} \\
$^{(e)}$ { Values taken from \citet{Aihara_19} and \citet{McCracken_12}. For the IRAC maps we employ the PSF publicly available on the IRSA website.} \\
$^{(f)}$ Two values for the deep/ultra-deep stripes in the UltraVISTA survey
\end{deluxetable*}

The analysis of DSFGs requires the most comprehensive wavelength coverage to account for all the physical processes taking place in these complex objects (i.e. stellar emission, dust obscuration,  thermal emission, and non-thermal processes). For this purpose, we extract accurate photometry in the optical-to-MIR regimes by analyzing the following maps:
\begin{itemize}
\item \textbf{Optical}: We analyze the maps produced with the Subaru telescope's Hyper Suprime Cam (HSC) during the Subaru Strategic Program ({ DR 3; \citealt{Aihara_19}), targeting the COSMOS field in the $g,r,i,z$ and $y$ bands}.
\item \textbf{NIR}: The photometry at NIR wavelengths is extracted from the DR4 maps of the UltraVISTA survey \citep{McCracken_12} performed with the VISTA telescope in the $Y, J, H$ and $Ks$ bands.
\item \textbf{MIR}: We analyze the maps produced with the \textit{Infrared Array Camera} (IRAC) of the \textit{Spitzer} space telescope. The maps analyzed in this study are the deepest ones made with this facility as part of the Cosmic Dawn Survey \citep{Moneti_22}, obtained by co-adding all the available exposures of the COSMOS field for each of the four channels of the IRAC camera.
\end{itemize}

Further details on the maps employed in the photometry extraction can be found in \Tab\ref{tab:maps}.

\subsection{Additional photometry}
\label{sec:additional}

To analyze { the dust emission in the FIR/(sub)mm and the non-thermal processes emitting at radio frequencies}, we retrieve additional photometry for our sources by cross-matching our sample with other catalogs in the current literature:
\begin{itemize}
\item \textbf{FIR}: We cross-match our catalog with the 2020 version of the SuperDeblended catalog by \citet{Jin_18}. Since the procedure followed in building that catalog employs as positional prior the sources in the VLA-COSMOS 3GHz Large Project, it is possible to retrieve FIR photometry obtained with { Spitzer (24 $\mu$m; \citealt{LeFloch_09}), Herschel (100, 160, 200, 250, and 500 $\mu$m; \citealt{Lutz_11}), SCUBA2 (850 $\mu$m; \citealt{Cowie_17,Geach_17}), AzTEC (1.1 mm; \citealt{Aretxaga_11}), and MAMBO (1.2 mm; \citealt{Bertoldi_07}) for all the galaxies in our sample. Further details on the employed maps and on their depth can be found in \citet{Jin_18}}.

\item \textbf{(sub)mm}: We cross-match our sample with the catalog of the \textit{Automated Mining of the ALMA Archive in the COSMOS Field} (A3COSMOS) survey (v.20200310; \citealt{Liu_19}). This catalog contains (sub)mm continuum fluxes for all the sources in the COSMOS field observed with ALMA. Since the coverage of the A3COSMOS survey is not uniform, only a tiny percentage of our RS-NIRdark sources has a counterpart in this catalog { (32 galaxies within a matching radius of 1'', $\sim 10\%$; \citealt{Liu_19})}, with central bandwidth and depth strongly variable for different sources.
\item \textbf{Radio}: Finally, we retrieve additional radio photometry by cross-matching our catalog with the public catalog by \citet{Schinnerer_10} containing 1.4 GHz photometry obtained during the VLA-COSMOS survey { (77 galaxies; $\sim24\%$ of the full sample)}. We also include radio photometry at 1.28 GHz from the MIGHTEE Early Science Data Release{ (170 sources within a matching radius of 8''; $\sim 53\%$ of the sample; \citealt{Jarvis_16,Heywood_22})}.
\end{itemize}

\begin{figure*}
    \centering
    \includegraphics[width=\textwidth]{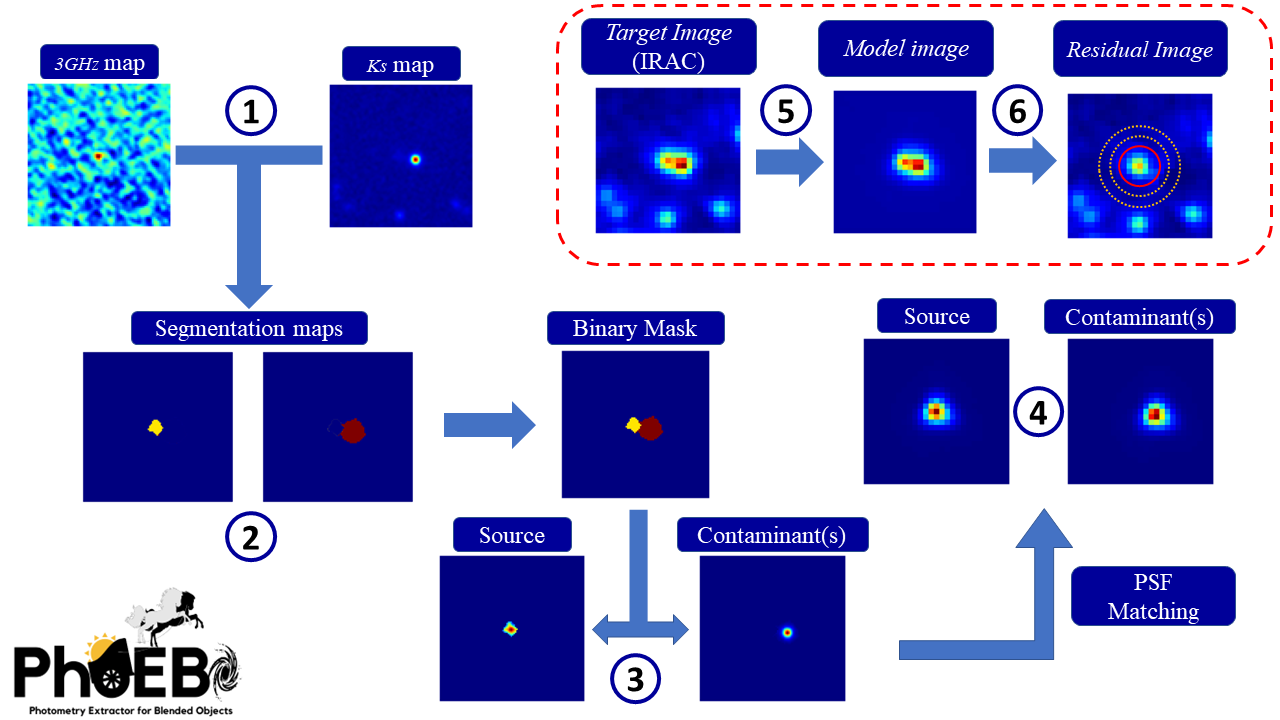}
    \caption{Scheme summarizing the workflow of the deblending algorithm \textsc{PhoEBO} employed to extract the optical/NIR/MIR photometry for the RS-NIRdark galaxies in the sample. Further details on the numbered steps and on the full procedure are given in \Sec\ref{sec:pipeline}. }
    \label{fig:photometry}
\end{figure*}

\section{\textsc{PhoEBO}: A new pipeline for photometry extraction}
\label{sec:photometry_extraction}

\subsection{Description of the pipeline}
\label{sec:pipeline}
Extracting accurate photometry for the sources in our catalog is a challenging task. The main limitation is represented by the possible presence of bright contaminants close to our (extremely faint or even undetected) galaxies in the optical/NIR bands. This issue becomes highly significant in the IRAC channels, where the large (up to FWHM$=2''$ in ch3 and ch4) and irregular Point Spread Function (PSF) makes it almost impossible to blindly deblend multiple sources without priors on their positions and shapes.

Basic deblending algorithms such as that implemented in \textsc{sExtractor} \citep{Bertin_96} need a minimum contrast between the different blended components inside the same cluster of pixel to recognize the presence of multiple sources. This level of contrast is not reached in the IRAC bands, making these algorithms unfit to analyze our galaxies. { More complex softwares such as \textsc{Tractor} \citep{Lang_16} and \textsc{Farmer} \citep{Weaver_22}, relying on profile-fitting, can overcome this problem by extracting prior information on the position and shape of the different objects through a high-resolution image in which all the blended components are present and distinguishable at the same time (e.g. the $\chi^2$-map employed by \citealt{Weaver_22}). As before, these techniques cannot be applied to our RS-NIRdark galaxies, since the only bands in which our galaxies are robustly detected - together with the contaminants - are the IRAC channels, where it is generally impossible to distinguish the different sources.} Therefore, in order to build the photometric catalog, we developed a new deblending pipeline called \textsc{PhoEBO} (Photometry Extractor for Blended Objects; \citealt{Gentile_PhoEBO})\footnote{The code is freely available in a \textsc{GitHub} repository: \url{https://github.com/fab-gentile/PhoEBO}} relying on a slightly modified implementation of the method employed in several previous studies \citep[e.g.][]{Labbe_06,Endsley_21,Whitler_22}. The whole procedure is summarized in \Fig\ref{fig:photometry} and follows these steps:
\begin{enumerate}
\item We start from two high-resolution “\textit{detection images}" and a low-resolution “\textit{target image}" to be deblended. One \textit{detection image} must contain information on the RS-NIRdark galaxy we want to analyze, the other on the contaminant sources. In our case, we choose the 3GHz and the $Ks$ maps, respectively. We underline that we minimize the risk of biases by employing two maps with a comparable PSF FWHM (0.75" and 0.78", respectively).
\item We employ the basic deblending algorithm included in the \textsc{Sep} library \citep{Bertin_96,Barbary_16a} to identify the different sources in the two \textit{detection images} and to produce the segmentation maps (i.e. a set of images assigning each pixel to one of the galaxies in the field). To include all the galaxies in the maps, we set a low detection threshold (2$\sigma$) and a minimum of 5 contiguous pixels in each detected source.
\item We combine the two segmentation maps and use their product as a binary mask to isolate the sources present in the \textit{detection images} and produce a different image for each of them.
\item We convolve each image produced in the previous step with a matching kernel to homogenize the PSF with that of the \textit{target image}. The matching kernel is produced with the “\textsc{photutils}” library \citep{Bradley_20} using the ratio of Fourier transforms \citep{Gordon_08,Aniano_11}. The PSF of the \textit{detection images} is modeled as a 2D Gaussian with a FWHM equal to the PSF FWHM of the $Ks$ band reported in \citet{McCracken_12} and \citet{Weaver_22}. The PSF of the optical/NIR bands are obtained in the same way {(i.e. with Gaussians with fixed widths, see \Tab\ref{tab:maps})}, while the PSFs of the four IRAC channels are downloaded from the IRSA archive\footnote{\url{https://irsa.ipac.caltech.edu/data/SPITZER/docs/irac/}}
\item We normalize all the resulting images and co-add them into a “\textit{model image}”. This one is resampled to match the pixel size of the \textit{target image} (if needed) and then fitted to the \textit{target image} by multiplying each normalized source by a free-parameter $\alpha_i$. The fit is performed with the \textsc{Scipy} library \citep{Virtanen_20}, aiming to minimize the $\chi^2$ between the \textit{model} and the \textit{target image} .
\item We multiply all the components of the \textit{model image} for the relative $\alpha_i$ obtained through the fitting procedure. Then, we subtract all the resulting images of the contaminants from the original \textit{target image}. In doing so, we get a \textit{residual image} containing only the source present in our sample.
\item Finally, we perform aperture photometry with \textsc{Photutils} on this residual image by employing a fixed diameter of 2 arcsec. The local background is computed in an annulus { between 1.5 and 2 times the radius of the aperture} and subtracted from the extracted counts. The counts are then converted to AB magnitudes and to micro-Jansky through the photometric zeropoints employed in \citet{Weaver_22}, corrected for the systematic offset reported in \citet{Weaver_22}. This last correction is needed to account for the systematic mismatch between the spec-\textit{z}s and the photo-\textit{z}s computed in the COSMOS field \citep{Laigle_16,Weaver_22}.
\end{enumerate}

The whole procedure described above allows us to estimate the fluxes in all the NIR and IRAC bands reported in \Sec\ref{sec:maps}. As prescribed by the IRAC Instrument Handbook, we correct the fluxes measured through aperture photometry by the factors reported in the IRAC Instrument Handbook\footnote{\url{https://irsa.ipac.caltech.edu/data/SPITZER/docs/irac/iracinstrumenthandbook/}}. { For the optical bands - with a smaller PSF FWHM than the radio map at 3GHz and the \textit{Ks} map - we subtract the contaminants with \textsc{PhoEBO} by using the HSC-\textit{i} band as a \textit{detection image}}. Finally, we estimate the photometric uncertainties as:
\begin{equation}
\label{eq:uncertainty}
    \Delta F = \sqrt{\sum_{i\in A} \sigma_i^2}
\end{equation}
where the $\sigma$ are obtained from the weight maps and the sum is extended to all the pixels in the aperture employed in the estimation of the flux. As prescribed by \citet{Weaver_22}, we correct the photometric uncertainties in the optical/NIR bands with the multiplicative factors reported in \Tab\ref{tab:maps}. This step is required to account for the expected under-estimation of the uncertainties through \Eq\ref{eq:uncertainty} due to the presence of correlated noise in the analyzed maps \citep[e.g.][]{Leauthaud_07}. { Since the weight maps are not affected by the subtraction of the contaminants, we underline that the procedure followed here to estimate the photometric uncertainties is totally consistent with that employed by \citet{Weaver_22} for the optical/NIR bands. Regarding the IRAC bands, we can compare our uncertainties with those by \citet{Weaver_22} by running \textsc{PhoEBO} on the 153 RS-NIRdark galaxies selected by \citet{Talia_21} and excluded from this study being revealed in the COSMOS2020 (see \Sec\ref{sec:selection}). We obtain that the median ratio between our uncertainties and those included in the “Classic" COSMOS2020 is in the order of $\sim3$. This result can be explained by the likely underestimation of the IRAC uncertainties found by \citet{Weaver_22} in the COSMOS2020 catalog.}

\subsection{Validation of \textsc{PhoEBO}}
\label{sec:validation}

\begin{figure}
    \centering
    \includegraphics[width=\columnwidth]{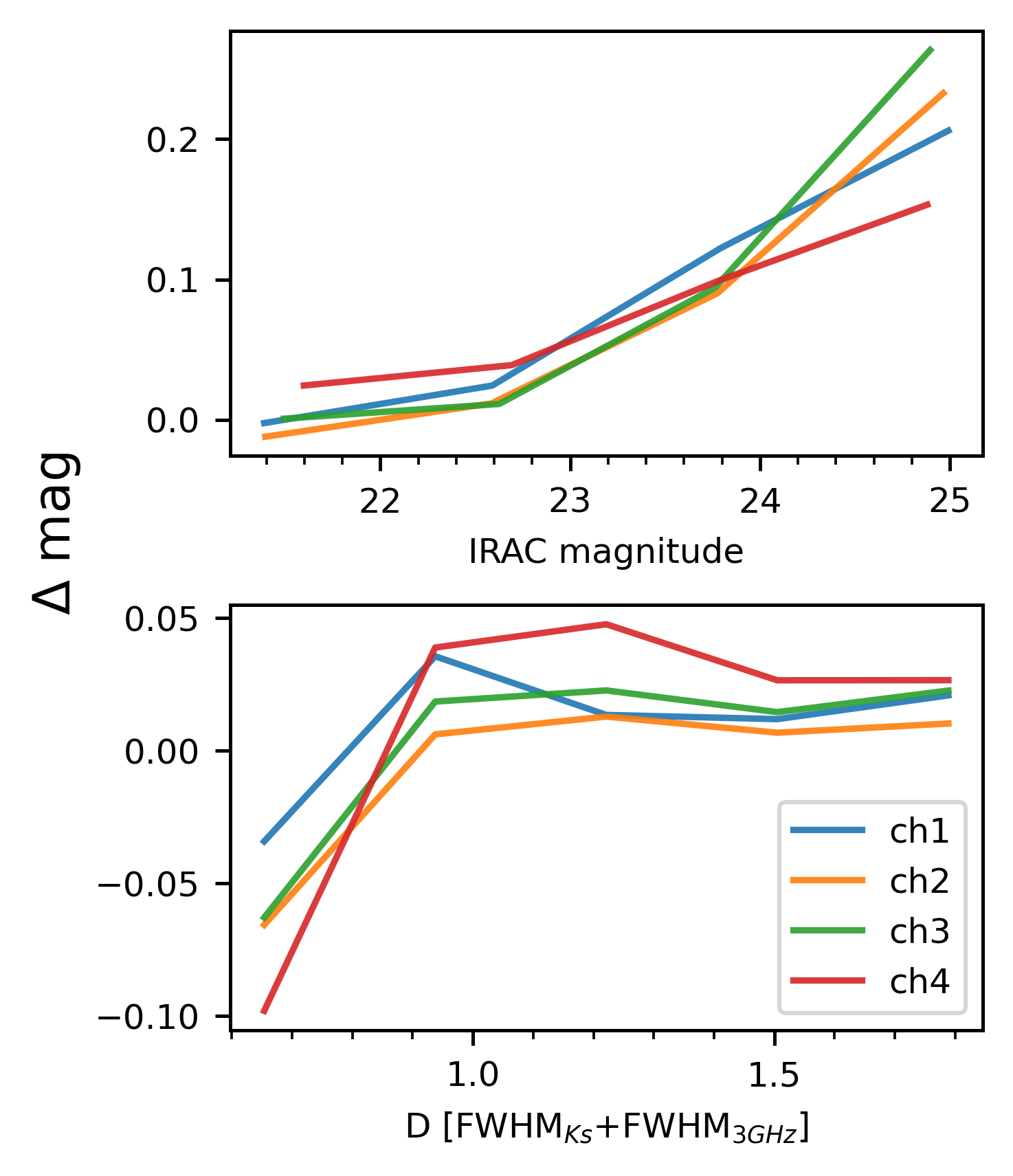}
    \caption{Accuracy of the \textsc{PhoEBO} pipeline as a function of the IRAC flux of the RS-NIRdark galaxies and of the angular distance between these sources and the blended contaminant (a proxy for the blending between the sources, once parameterized through the sum of the FWHMs of the two sources). These results are obtained by applying the \textsc{PhoEBO} pipeline to the set of simulated images described in \Sec\ref{sec:validation}.}
    \label{fig:Validation}
\end{figure}

Thanks to the small difference in the PSFs, the photometry extraction performed by \textsc{PhoEBO} in the optical/NIR bands does not differ significantly from that performed by “classic" algorithms such as \textsc{sExtractor} \citep{Bertin_96}. Nevertheless, the extraction in the IRAC channels is quite different. Therefore, in order to obtain reliable results on the physical properties extracted from the photometric catalog, we need to validate the performances of the pipeline in the MIR regime. We validate the results of the \textsc{PhoEBO} pipeline performing extensive simulations of blended galaxies in the four IRAC channels. The simulation procedure recalls the philosophy discussed in \Sec\ref{sec:pipeline}:
\begin{enumerate}
    \item We simulate two noise maps through a random Gaussian generator, requiring an rms compatible with the 3GHz images observed in the VLA-COSMOS survey at 3 GHz and with the $Ks$ band images observed in the UltraVISTA survey (conservatively, we employ the sensitivity reached in the deep stripes).
    \item  We simulate a radio source and a NIR-bright contaminant. Both the galaxies are simulated as 2D Gaussians on the noise maps generated in the previous step. Since the vast majority of the RS-NIRdark galaxies are unresolved at 3 GHz, we choose a FWHM=0.7" for the radio sources (i.e. the width of the synthesized beam in the VLA-COSMOS survey). On the contrary, to account for the presence of partially resolved contaminants, the FWHMs of the Gaussians in the $Ks$ band are uniformly sampled in the range [0.7,1.4] arcsec (i.e. between one and two PSF FWHM). The normalizations of the two Gaussians are chosen to obtain a S/N uniformly sampled in the range [5.5,8] and [3,8], for the radio and $Ks$ images, respectively. The radio sources are placed in the center of each image, while the contaminants are allowed to space in the range [0.7,1.4]arcsec from the center. This range allows us to study the accuracy of the algorithm as a function of the blending between the two sources. The lower limit on the distance is chosen as 0.7'', since we recall that - due to the selection described in \Sec\ref{sec:selection} - all the NIR sources with a separation lower than this threshold are considered NIR counterparts of the radio signal.
    \item We convolve each of these Gaussians with a set of matching kernels computed as prescribed in \Sec\ref{sec:pipeline} to obtain the sources as observed in each of the four IRAC channels.
    \item We rescale the flux of the convolved radio source to obtain an integrated flux in the range [21,M] mag, where M is the limiting magnitude in each IRAC channel. Consequently, we rescale the flux of the convolved contaminant to obtain a flux ratio uniformly sampled in the range [0.1,1] (with the radio source brighter than the other galaxy).
    \item Finally, we co-add the two images and add Gaussian noise with the same rms expected in the four IRAC channels.
\end{enumerate}

\begin{figure*}
    \centering
    \includegraphics[width=\textwidth]{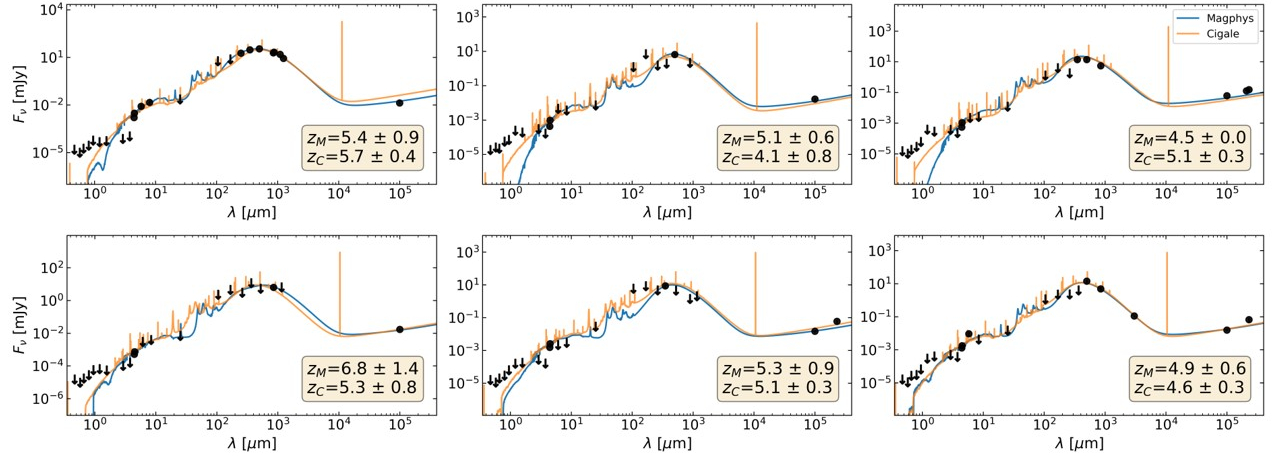}
    \caption{Some examples of SEDs of high-\textit{z} candidates fitted with \textsc{Magphys} (blue solid lines; \citealt{daCunha_08,Battisti_19}) and \textsc{Cigale} (orange solid lines; \citealt{Boquien_19}). The photometry is represented by the black dots (for the detections at $S/N>3$) and the black arrows (for the upper limits). The yellow boxes report the photo-\textit{z}s computed with the two codes and the relative uncertainties.} 
    \label{fig:SED}
\end{figure*}

To explore the whole parameter space of the randomly sampled quantities, we simulate $\sim10^3$ images. Then, we run the \textsc{PhoEBO} pipeline on the IRAC-like images, employing as \textit{detection images} the high-resolution data in the radio and $Ks$ bands. In order to assess the performances of the pipeline, we compare the fluxes reported by \textsc{PhoEBO} with those obtained by performing a standard aperture photometry on the isolated IRAC-like images (i.e. those obtained in point 3 of the simulation procedure, before adding the contaminant).

Computing the $\Delta$mag on the whole dataset, we obtain a median($\Delta$mag)$\sim$0.03 in all the IRAC channels, without any significant difference between the different bands. Similarly, we obtain a std($\Delta$mag)$\sim$0.15 in all the channels. Moreover, the employment of simulations allows us to estimate the $\Delta$mag as a function of the different parameters employed during the simulation procedure (\Fig\ref{fig:Validation}). We find interesting - albeit expected - trends with the IRAC flux and with the angular separation between the high-resolution images, with lower accuracy achieved on more blended sources and IRAC-fainter RS-NIRdark galaxies.

\section{Physical properties from SED-Fitting}
\label{sec:properties}

To determine the nature of the RS-NIRdark galaxies, we need to assess their photo-\textit{z}s and physical properties. We do this through an SED-fitting procedure. To test the robustness of our results against the characteristics of different codes, we base our analysis on two algorithms: \textsc{Magphys+photo-\textit{z}} \citep{daCunha_08,Battisti_19} and \textsc{Cigale} \citep{Boquien_19}.

\begin{figure*}
    \centering
    \includegraphics[width=\textwidth]{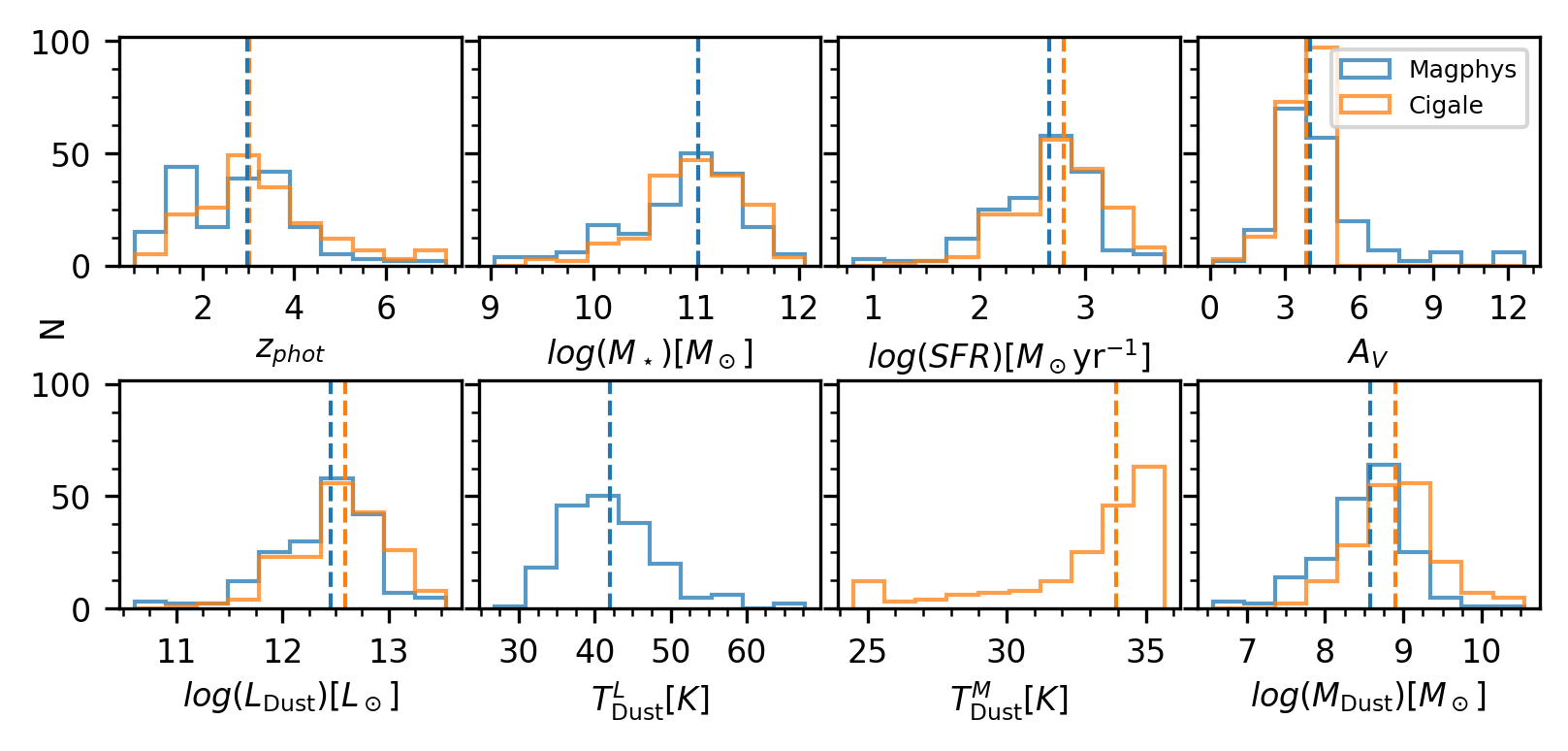}
    \caption{Distribution of the photo-\textit{z}s and of the main physical properties { (in order: stellar mass, star formation rate, extinction, dust luminosity, dust temperature - both luminosity- and mass-weighted - and dust mass)} of the RS-NIRdark as estimated by SED-fitting with \textsc{Magphys+photo-z} \citep{daCunha_15,Battisti_19} and \textsc{Cigale} \citep{Boquien_19} in blue and orange, respectively. The dashed lines of the same colors mark the median values of the distributions. Further details are given in \Sec\ref{sec:properties}.}
    \label{fig:Properties}
\end{figure*}

\subsection{SED-fitting with \textsc{Magphys}}
\label{sec:magphys}

\textsc{Magphys} \citep{daCunha_08} is a physically-motivated SED-fitting code based on the energy balance between stellar attenuation and thermal dust emission. The software estimates the physical properties of a galaxy by comparing its optical-to-radio broad-band photometry with more than a million templates, including stellar emission, dust attenuation, thermal dust emission and non-thermal radio emission. 
The stellar emission is considered by combining the Single Stellar Populations (SSPs) by \citet{Bruzual_03} with an exponentially declining Star Formation History (SFH) with random bursts of star formation superimposed on the continuum. The dust attenuation is included as prescribed by \citet{Charlot_00}, with the addition of a $2175A$ feature accounting for the young stars born in denser clouds. A three-component model accounts for thermal dust emission. These components include the “hot dust” (i.e. warmed up by young stars in the birth clouds), the “cold dust” (i.e., present in the diffuse interstellar medium), and the characteristic emission by the Polycyclic Aromatic Hydrocarbures (PAHs). The first two components are modeled with a modified gray body, while the PAH emission is modeled with an empirical template (see \citealt{daCunha_08} for details). { The code computes the dust temperature ($T_{\rm d}$) as the luminosity-average of these three components.} Finally, \textsc{Magphys} includes the radio emission from star formation as prescribed by \citet{daCunha_15}. All the ranges of the free parameters employed in the templates can be found in \citet{daCunha_08} and \citet{daCunha_15}. In this work, we use the \textsc{photo-\textit{z}} version of \textsc{Magphys} \citep{Battisti_19}, able to estimate the photometric redshift together with the physical properties of the analyzed galaxies.

\subsection{SED-fitting with \textsc{Cigale}}
\label{sec:cigale}

To avoid possible biases arising from the use of a single SED-fitting code, we involve the software \textsc{Cigale} \citep{Boquien_19} in the analysis. This code is based on a similar energy-balance principle as \textsc{Magphys} but allows a larger customization of the libraries employed in building the templates. In this paper, we start by using a setup as close as possible to \textsc{Magphys} to achieve consistent results. A detailed investigation of the different modules included in \textsc{Cigale} will be discussed in a forthcoming paper. The SSPs, SFH, dust attenuation and radio emission are the same as discussed in \Sec\ref{sec:magphys}. A significant difference between the two codes resides in the treatment of the thermal dust emission. \textsc{Cigale} does not have a single model including both the gray-body thermal emission and the PAH emission as \textsc{Magphys}. The choice is limited to the models by \citealt{Draine_07} and their updated version by \citealt{Draine_14} parameterized through the intensity of the radiation field and including the PAHs emission. A second possibility is the analytical model by \citet{Casey_12} parameterized through the dust temperature but not including the PAHs. Since our photometric catalog includes a point at 24$\mu$m from the SuperDeblended catalog, sampling the typical PAHs emission at $z\sim3$ and since this feature is generally crucial for determining a robust redshift, we decide to employ the \citet{Draine_14} model. { We compute the mass-weighted dust temperature starting from the intensity of the radiation field reported in output by \textsc{Cigale} following the framework described in \citet{Draine_07} and \citet{Draine_14}.}

\begin{deluxetable*}{cccccc}
\label{tab:properties}
\tablecaption{Comparison between the median properties estimated by \textsc{Magphys} and \textsc{Cigale}}
\tablewidth{0pt}
\tablehead{
\colhead{Property} & \multicolumn{2}{c}{\textsc{Magphys}} &\multicolumn{2}{c}{ \textsc{Cigale}} & \colhead{Unit} \\
 & \colhead{Median} & \colhead{$\sigma$} & \colhead{Median} & \colhead{$\sigma$} & 
}
\startdata
$z_{\rm phot}$ & $2.96\pm 0.04$ & 1.2 & $3.02\pm 0.04$& 1.3 & \\
$\log(M_\star)$ & $11.01 \pm 0.02$ & 0.61 &$11.01 \pm 0.02$ & 0.44 &$M_\odot$ \\
$\log(SFR)$ &$2.67 \pm 0.02$ & 0.48 &$2.79 \pm 0.04$& 0.45 &$M_\odot {\rm yr}^{-1}$ \\
$A_v$ & $4.01\pm0.04$& 1.3 &$3.89\pm0.05$ & 0.31& mag\\
$\log(L_{\rm Dust})$ & $12.45\pm0.02$ & 0.48& $12.59\pm0.02$ & 0.44& $L_\odot$ \\
$T^L_{\rm Dust}$ &  $42.05\pm0.2$ & 5.8 & -- && K\\
$T^M_{\rm Dust}$ &  -- & &$33.9\pm0.2$ & 2.7& K\\
$\log(M_{\rm Dust})$&$8.57 \pm 0.02$& 0.5 &$8.9 \pm 0.02$ & 0.46 &$M_\odot$
\enddata 
\textbf{Note:}\\
The uncertainties on the median properties are estimated as the Median Absolute Deviation: $MAD=1.482 \times {\rm median}(|x_i-{\rm median}(x_i)|)$ \citep{Hoaglin_83} divided by $\sqrt{N}$, where N is the number of galaxies in the sample. { For each quantity, we also report the dispersion computed as half the symmetrized interval between the 16th and the 84th percentiles.}
\end{deluxetable*}

\subsection{SED-fitting results and comparison between the codes}

The two codes are applied to the photometric catalog presented in \Sec\ref{sec:photometry_extraction}. { Before running the codes, we correct the photometry in the catalog for the galactic extinction. We employ the dust maps by \citet{Lenz_17} and the extinction law by \citet{Fitzpatrick_07}. This correction is performed through the python libraries \textsc{DustMaps} \citep{Green_18} and \textsc{Extinction} \citep{Barbary_16b}.} To account for the possible biases in the photometry extraction (see \Sec\ref{sec:validation}), to account for the known under-estimation of the photometric uncertainties by the \textsc{Photutils} library employed in \textsc{PhoEBO} \citep[see e.g][]{Leauthaud_07,Laigle_16,Weaver_22} and to allow the SED-fitting codes to explore a wider region in the photometry space, we add in quadrature 0.15 mag to the photometric uncertainties included in the catalog for the optical/NIR/MIR bands, following \citet{Laigle_16} and \citet{Weaver_22}. 
The output of the SED-fitting codes \textsc{Magphys} and \textsc{Cigale} are summarized in \Fig\ref{fig:Properties}, while the median values of the photo-\textit{z}s and the main physical properties for both codes are reported in \Tab\ref{tab:properties}. { In generating the histograms and in computing the medians, we exclude from the sample 57 sources that could host an AGN (see the discussion in \Sec\ref{sec:AGN}), in order to avoid possible biases coming from the SED-fitting performed with templates not including nuclear activity. { Similarly, we do not include in our analysis 10 sources with no other robust detections than in the radio (the so-called “type 0" of \citealt{Behiri_23}, see that study for the possible models of these sources) for which the properties estimated through SED-fitting would be highly unreliable.} The good convergence of the SED-fitting procedure is ensured by the median reduced $\chi^2<2.5$ obtained by both codes on the analyzed sample. Since for our galaxies the constraints on stellar population in the optical/NIR regimes are limited - in the majority of the cases - to upper limits, we compute the SFR starting from the infrared luminosity through the relation by \citet{Kennicutt_12} rescaled to a \citet{Chabrier_03} IMF. { This quantity is more robustly estimated, since $\sim$ 60\% of the galaxies in the sample have a detection at FIR/(sub)mm wavelengths. { However, we must underline how this quantity is more unconstrained for the other $\sim$40\% of the sources in the sample without these detections. Nevertheless, since the two codes employed in the SED-fitting procedure rely on the energy balance and include the radio fluxes, some constraints on the infrared luminosity can also come by the modeled dust attenuation and by the radio luminosity \citep{daCunha_15,Battisti_19,Boquien_19}. Finally, we notice that no significant discrepancy is visible when comparing the SFR estimated through the $L_{\rm IR}$ and the $L_{\rm 1.4 GHz}$ (see e.g. \citealt{Kennicutt_12,Novak_17}) for the galaxies with FIR/(sub)mm detections and those without. This result can be partly explained by the fact that \textsc{Magphys} assumes a constant radio-infrared correlation with a $q_{\rm TIR}$ centered on 2.34 and with a $1\sigma$ dispersion of 0.25 \citep{daCunha_15,Battisti_19}. This value is broadly compatible with the median $q_{TIR}=2.1\pm0.3$ computed for the galaxies in our sample (see \Sec\ref{sec:qtir}).}

As it can be seen, the results of the two software for what concerns the photo-\textit{z}s, and most of the physical properties are broadly compatible. The slight difference between the outputs for these quantities can be explained by some minor differences between the codes (e.g. the treatment of the upper limits, representing the majority of the constraints in the optical/NIR regime for our galaxies; see e.g. \citealt{Battisti_19} and \citealt{Boquien_19}). Major differences hold for the dust mass and temperature. As discussed in \Secs\ref{sec:magphys} and \ref{sec:cigale}, \textsc{Magphys} and \textsc{Cigale} report the luminosity- ($T^L_{\rm Dust}$) and mass-weighted ($T^M_{\rm Dust}$) dust temperature, respectively. Since these quantities weight differently the hot and cold components of the dust \citep[see e.g.][]{Liang_19,Sommovigo_20}, it is not surprising the difference in the two distributions reported in \Fig\ref{fig:Properties}. Moreover, this difference in the two codes can also explain the discrepancy in the two estimates of the dust mass, being this quantity computed starting from the dust temperature.} { In the following, for consistency with the previous papers of the series \citep[]{Talia_21,Behiri_23}, we will consider the results obtained by \textsc{Magphys}.} { A final note concerns the accuracy of our photo-\textit{z}s. Unfortunately, due to the elusive nature of the sources in our sample, it is not easy to retrieve spectroscopic redshifts for our RS-NIRdark galaxies from the current literature. One spectroscopic redshift can be found in \citet{Jin_22}, where they targeted one of our galaxies with a spectral scan performed with the ALMA and NOEMA interferometers. For our galaxy RSN-436 (ID 3117 in \citealt{Jin_22}), they obtained $z_{\rm spec}=3.545$, which is in good agreement with our estimates ($z_{\rm M}=3.3\pm0.3$ and $z_{\rm C}=3.4\pm0.2$. To increase the spectroscopic coverage of our sample, we already have planned approved spectroscopic follow-ups with (sub)mm facilities such as ALMA and NOEMA (Gentile et al., subm.).}

\section{AGN contribution}
\label{sec:AGN}

While in \Sec\ref{sec:intro} we discussed the possible biases affecting the FIR and (sub)mm selection of DSFGs, this section focuses on the main bias possibly affecting our radio selection: the presence of strong Active Galactic Nuclei (AGN). Since both star formation and radio activity can generate radio emission, we cannot exclude that some of our galaxies are hosting an AGN. For instance, \citet{Bonzini_13, Novak_17} and \citet{Enia_22}, analyzing samples of radio-bright galaxies selected at 1.4 GHz, reported significant fractions of AGN in their catalogs, spanning from 10\% (for the faintest radio fluxes) to 100\% (for the radio-brightest sources). Fortunately, our selection procedure focuses on galaxies without a significant optical/NIR counterpart. This step allows us to remove from the sample all the brightest AGN. This selection, however, does not allow us to remove from the sample the obscured AGN, where a significant amount of dust absorbs the optical/NIR emission (see e.g. the review by \citealt{Hickox_18} and references therein). The presence of these sources in our sample can be unveiled by searching for AGN tracers at longer wavelengths (namely the IR and radio regimes), or in the X-ray, taking advantage of the broad wavelength coverage of our sample. 
Estimating the fraction of obscured AGN and properly accounting for their presence is crucial for two main reasons. Firstly, the presence of an extra IR component due to a dusty torus surrounding the AGN and/or a radio excess due to nuclear activity can bias the SFR when obtained from the IR/radio luminosity through the relation by \citet{Kennicutt_12}. { Similarly, the employment of galaxy-only templates could bias all the other physical properties (and the photo-\textit{z}s) estimated through SED-fitting}. Secondly, the contamination affecting the selection procedure has to be considered in the determination of the statistical properties of the sample (e.g. the luminosity function and the contribution to the SFRD).

\begin{figure*}
    \centering
    \includegraphics[width=\textwidth]{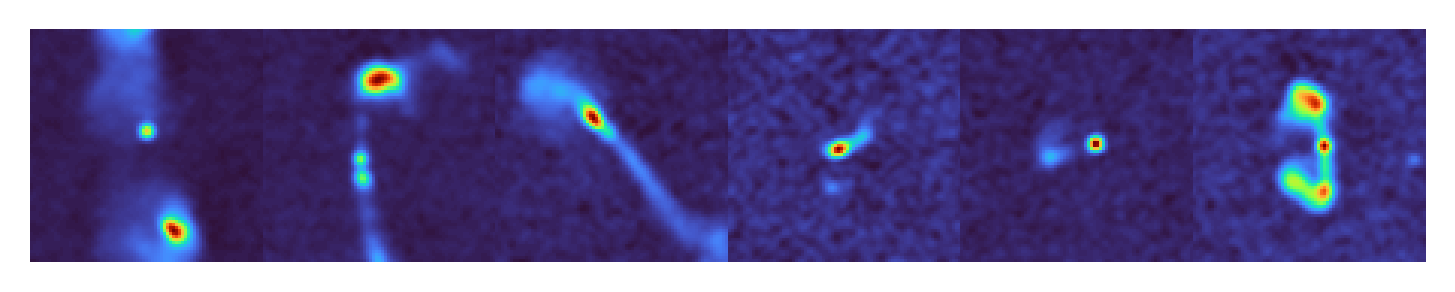}
    \caption{Some examples of the sources marked as likely AGN in the final catalog because of their morphology during the visual inspection of the radio maps at 3GHz. The postage have a 15 arcsec side. Further details are given in \Sec\ref{sec:AGN}.}
    \label{fig:visual_AGN}
\end{figure*}

\subsection{Visual inspection}
The first selection of galaxies hosting AGN in our sample is performed through a visual inspection of the 3GHz radio maps. This procedure aims to select all the galaxies with radio morphologies generally associated with AGN (e.g. radio blobs and relativistic jets). This visual inspection allows us to mark as likely AGN 17 galaxies. Some examples of these sources are reported in \Fig\ref{fig:visual_AGN}. We underline that - although this selection is 100\% pure (since non-active galaxies do not have these peculiar radio morphologies) - it is rather far from complete since most of the galaxies in the sample are unresolved at the 0.7” resolution of the 3GHz maps by \citealt{Smolcic_17}. Moreover, 9 RS-NIRdark galaxies are also part of the catalog by \citet{Verdoulaki_21} containing AGN in the COSMOS field selected through visual inspection thanks to their morphology in the same radio maps employed for our selection. Cross-matching our sample with this catalog, we find out that all the 9 sources are classified by AGN by both the selections. This results strengthen the reliability of our classification.

\subsection{X-ray stacking}
\label{sec:xray}

Another test to unveil the presence of AGN in our sample can be performed in the X-ray regime. We conduct two complementary tests to characterize the sample of RS-NIRdark galaxies as a whole and the individual sources.

\begin{itemize}
\item \textbf{Sample characterization:} The first test is based on the median X-ray flux computed on all the galaxies in our sample and on the possibility of explaining it by only invoking the star formation. We follow this procedure:

    1. We extract the X-ray flux for each RS-NIRdark galaxy through aperture photometry on the { event file in the range [0.5,7] keV} produced by the Chandra telescope in the C-COSMOS \citep{Elvis_09} and COSMOS legacy \citep{Civano_16} surveys. We perform the extraction through the \textsc{dmextract} function of the \textsc{CIAO} library \citep{Ciao_06}, employing circular apertures centered on the radio position of each source. We choose the radius of each aperture by assuming that each source is not resolved (a reasonable hypothesis for possible AGN) and employing the radius used in \citet{Elvis_09} and \citet{Civano_16} to extract the flux of the X-ray source closest to the considered galaxy. More in detail, we employ the median radius (considering all the observations in which the source fell) corresponding to an area encompassing 90\% of the PSF at the X-ray closest source. The local background is computed and subtracted from each galaxy through an annulus with an outer radius 1.5 times the circular aperture. 
    
    2. Once the net counts are obtained, these are converted to X-ray luminosities through the \textsc{PIMMS}\footnote{https://heasarc.gsfc.nasa.gov/docs/software/tools/pimms.html} software by assuming a power-law model with a slope of $\Gamma=1.8$ (i.e. that expected for possible AGN) and a galactic $n_H=1.7\times10^{20}$ cm$^{-2}$ \citep{Civano_16}. 
    
    3. Finally, we convert the luminosities in fluxes by assuming the photometric redshifts computed by \textsc{Magphys} (\Sec\ref{sec:properties}) and these in SFR through the empiric relation by \citet{Ranalli_03}. 

Considering the median SFR obtained by these X-ray fluxes, we obtain a value of $\log(SFR_X)=(2.24\pm0.02)M_\odot {\rm yr}^{-1}$, slightly lower than that obtained through the FIR flux ($\log(SFR_{IR})=(2.38\pm0.02){\rm M}_{\odot}{\rm yr}^{-1}$. This result confirms the lack of strong un-obscured AGN activity in our sample.

\item\textbf{Source characterization:} If the analysis performed in the previous paragraph ensures that the bulk of the sample of RS-NIRdark is composed of SFGs, it does not give insights on the presence of single AGN within the total sample. To address this point, we perform an additional analysis. We cross-match the two catalogs of X-ray sources in the COSMOS field by \citet{Elvis_09} and \citet{Civano_16} with our sample by employing as matching radii the positional uncertainty included in these catalogs (for the X-ray sources) and 0.7” (for the NIRdark galaxies). { Given the relatively shallow depth of the X-ray coverage in the COSMOS field, all these sources have an X-ray luminosity higher than $10^{42}$ erg s$^{-1}$, therefore, we can assume that their X-ray flux is largely due to the presence of an AGN (see e.g. \citealt{Hickox_18}).} Once the list of RS-NIRdark galaxies with a possible X-ray counterpart is obtained, we visually inspect the NIR and radio maps at 3GHz. We obtain two main cases:

1. The positional uncertainty on the X-ray source includes only the RS-NIRdark galaxy. In this case, we can safely assume that the X-ray signal is produced by the galaxy in our sample, suggesting the presence of an AGN in that galaxy. The 3 sources in this class are marked in the photometric catalog with an appropriate flag.
    
2. The positional uncertainty of the X-ray source includes both an RS-NIRdark galaxy and a NIR-bright galaxy. In this case, we cannot unambiguously associate the X-ray signal to one of these galaxies. The { 15} NIR-dark galaxies in this class are marked in the catalog with a different flag accounting for the possibility of hosting an AGN.

\end{itemize}

\subsection{$q_{\rm TIR}$ analysis}
\label{sec:qtir}

A standard method to identify AGN relies on the so-called infrared-radio correlation. It is well established that the radio luminosity measured at 1.4 GHz and the infrared luminosity measured in the range [8;1000]$\mu$m are tightly ($\sigma \sim 0.16$ dex; e.g. \citealt{Molnar_21}) correlated in star-forming galaxies (see e.g. \citealt{deJong_85,Helou_85}). This correlation is generally measured through the $q_{\rm TIR}$ parameter defined as (see e.g. \citealt{Helou_85,Yun_01}):  
\begin{equation}
    q_{\rm TIR}=\log\left(\frac{L_{\rm IR} {[W]}}{3.75 \times 10^{12} {\rm Hz}} \right) - \log\left(\frac{L_{1.4 \rm GHz}}{{\rm W}\,{\rm Hz^{-1}}} \right)
\end{equation} 
and mainly arises because of the connection between the star formation, the radio synchrotron emission in star forming regions and the thermal FIR emission of dust in the same regions. Several authors studied the possible evolution of the $q_{\rm TIR}$ parameter with cosmic time or possible correlations with other physical quantities. In this work, we explore two of the main studies on this point. \citet{Delhaize_17} found a possible evolution with the redshift through the relation
\begin{equation}
\label{eq:Delhaize}
    q_{\rm TIR}(z)=(2.88\pm0.03)\times(1+z)^{(-0.19\pm0.01)} 
\end{equation}

On the contrary, more recently, \citet{Delvecchio_21} suggested a $q_{\rm TIR}$ almost constant with the redshift, but strongly dependent on the stellar mass of the hosting galaxy through the relation
\begin{multline}
\label{eq:Delvecchio}
    q_{\rm TIR}(M_\star,z)=(2.646 \pm 0.024)\times A^{(-0.023\pm0.008)}+ \\ -B\times(0.148\pm0.013) 
\end{multline}
where $A=(1+z)$ and $B=(\log(M/M_\odot)-10)$.

We compute the radio luminosity at 1.4 GHz for the galaxies with a counterpart in the 1.4 GHz catalog by \citet{Schinnerer_10} through the relation
\begin{equation}
    L_{\rm 1.4 GHz}=\frac{4 \pi D_L^2(z)}{(1+z)^{1+\alpha}}S_{\rm 1.4 GHz}
\end{equation}
by employing the photometric redshifts estimated by \textsc{Magphys} and the spectral slope $\alpha$ computed through the flux densities reported in the 3 GHz and 1.4 GHz catalogs. For the galaxies without a counterpart in the 1.4 GHz sample, we evaluate the 1.4 GHz flux starting from the 3GHz flux density through the relation
\begin{equation}
    L_{\rm 1.4 GHz}=\frac{4 \pi D_L^2(z)}{(1+z)^{1+\alpha}}\left(\frac{1.4 {\rm GHz}}{{\rm 3 GHz}}\right)^\alpha S_{\rm 3 GHz}
\end{equation}
and assuming a spectral slope of $\alpha=-0.7$ (generally considered for star-forming galaxies; see e.g. \citealt{Novak_17}). 

Once we applied \Eqs\ref{eq:Delhaize} and \ref{eq:Delvecchio} to our sample { (\Fig\ref{fig:qtir})}, marking as likely AGN the sources distant more than { 3$\sigma$ from the relations \citep[see e.g.][]{Delvecchio_17,Enia_22}, we obtain that 40 sources are classified as AGN following \citet{Delhaize_17} and 57 following \citet{Delvecchio_21}. We include the results from both the tests in the final catalog. However, given the high uncertainties affecting our photo-\textit{z}s and our stellar masses, in the following we consider as likely AGNs only the 37 sources classified as AGN according to both the relations. { A final interesting remark concerns the overall distribution of the $q_{\rm TIR}$ as a function of the photometric redshift. As visible in \Fig\ref{fig:qtir}, the bulk of the population of the RS-NIRdark galaxies have a median $q_{\rm TIR}$ compatible with that expected at the median redshift of the sample from the relation by \citet{Delhaize_17} ($2.1\pm0.3$ and $2.21\pm0.05$, respectively). However, the distribution of these values at low redshift appear to significantly differ from that expected from the relation. This discrepancy can be explained with the different selection operated in this study with respect to \citet{Delhaize_17}. In particular, our selection of NIR-dark sources is expected to produce a sample of more extremely-obscured sources at low-\textit{z} to account for the extinction of redder rest-frame bands. Therefore, we do not expect in this regime the properties of our (incomplete) sample to completely resemble those of the total population of radio-selected galaxies analysed in \citet{Delhaize_17}.}

\begin{figure}
    \centering
    \includegraphics[width=\columnwidth]{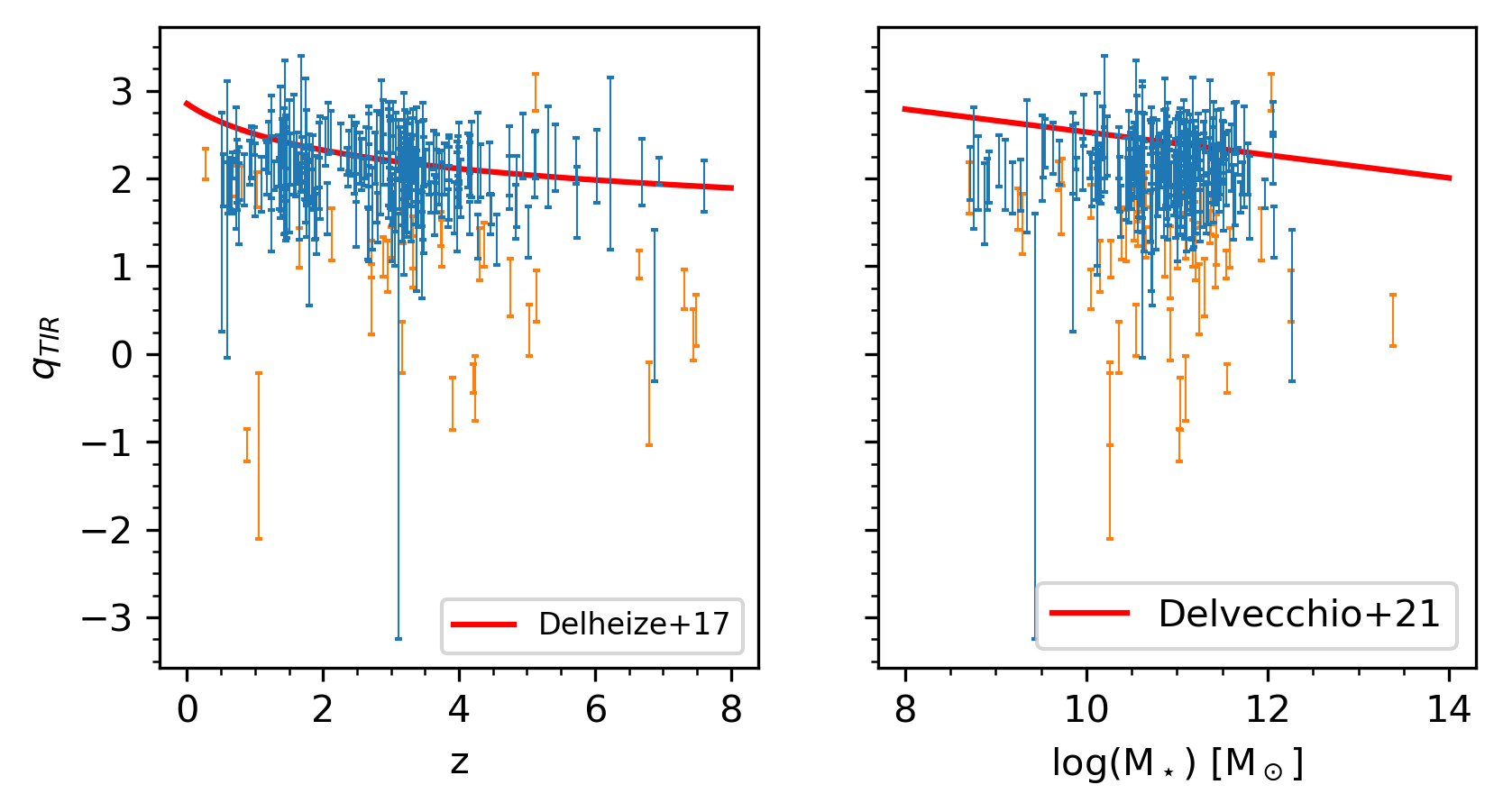}
    \caption{{ Behavior of the $q_{\rm TIR}$ as a function of the redshift (left panel) and stellar mass (right panel). The blue solid lines show the relation by \citet{Delhaize_17} and \citet{Delvecchio_21}, respectively. The objects distant more than $3\sigma$ from each relation are highlighted in red.}}
    \label{fig:qtir}
\end{figure}

\subsection{SED decomposition}

A further sign of AGN hosted in our galaxies can be found at MIR wavelengths. It arises from the presence of a dusty torus surrounding the supermassive black hole and heated up from the high-energy radiation coming from there. Trying to model the SED with a template not accounting for this additional component generally produces a best-fitting SED under-estimating all the MIR fluxes \citep[see e.g.][]{Hickox_18}. This problem can be solved by adding a torus component to the templates fitted to the galaxy photometry. We perform this test with \textsc{Cigale} (\textsc{Magphys} does not allow the addition of an AGN component). The overall setup is the same as discussed in the \Sec\ref{sec:cigale}, but adding the dusty torus component as modeled by \citet{Fritz_06}. The model’s parameters are the same as employed in \citet{Donevski_20} in modeling a sample of likely DSFGs. The main parameter describing the effect of the AGN on the modeled SED is the AGN fraction $f_{\rm AGN}$, defined as the ratio between the torus luminosity and the dust luminosity in the range $[5,40]\mu$m. Defining as likely AGN all the galaxies with a $f_{\rm AGN}>10\%$, we mark 11 sources among the entire sample of RS-NIRdark galaxies. We underline that the median value of $f_{\rm AGN}$ computed on the whole sample is compatible with zero, suggesting an overall small contribution of AGN in the RS-NIRdark galaxies. { However, we underline that - due to the limited coverage of the MIR regime in our sample - also this estimation should be considered as a lower limit on the actual AGN contribution in our galaxies.}

\subsection{Final remarks on AGN contamination}
The sample of likely AGN reported by the different methods have some overlap, as shown in \Fig\ref{fig:Venn}. Considering all the galaxies marked as possible AGN by at least one method, we obtain a sample of 64 sources ($\sim23\%$ of the full sample). Excluding the sources with an uncertain X-ray counterpart (i.e. those at point 2 in \Sec\ref{sec:xray}) and with no other indication of AGN activity from the other methods, we obtain 57 sources ($\sim 20\%$ of the sample). 

As a final remark, we can compare the estimated fraction of RS-NIRdark galaxies hosting an AGN with a theoretical prediction based on analogous works present in the literature. \citet{Bonzini_13} and \citet{Novak_18}, working on radio-selected catalogs of galaxies, reported in their work the fraction of AGN in the analysis of their sources as a function of the radio flux at 1.4 GHz. Considering these relations and integrating them on our\footnote{Here, we employ the radio fluxes at 1.4GHz provided by \citet{Schinnerer_10}. For the sources without a counterpart we conservatively assume a flux equal to the sensitivity of the survey} flux distribution at 1.4 GHz, we estimated an expected fraction of AGN in our sample lower than 28.3\% (further divided in 27\% of radio-quiet AGN and 1.3\% of radio-loud AGN). We underline that this last estimation represents an upper limit on the expected number of AGN in our sample, since we expect that focusing on radio sources without an optical/NIR counterpart contributes to exclude a significant fraction of these objects.

\begin{figure}
    \centering
    \includegraphics[width=\columnwidth]{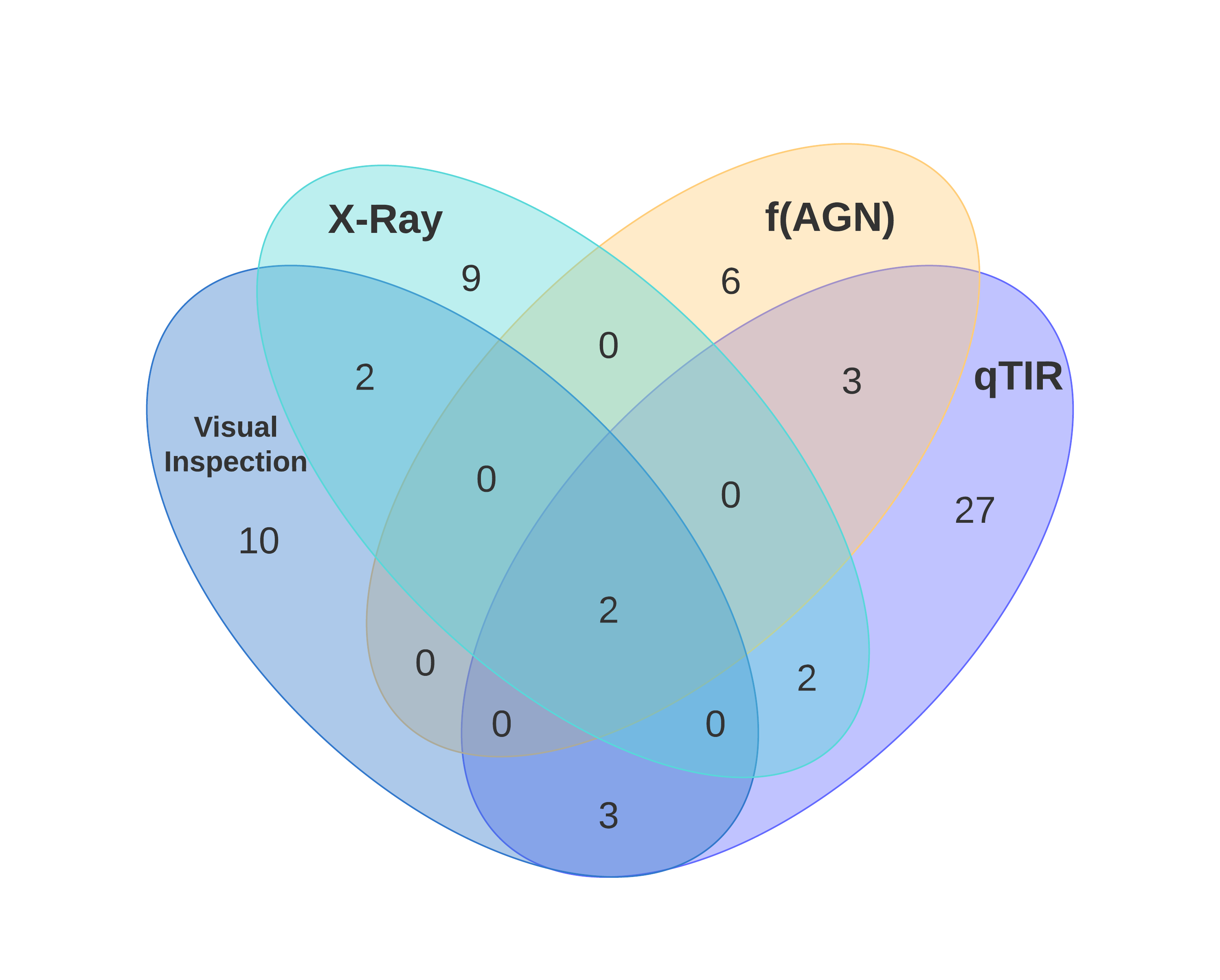}
    \caption{Diagram reporting the number of likely AGN unveiled by the different methods described in \Sec\ref{sec:AGN} and the relative overlap. Further details in \Sec\ref{sec:AGN}.}
    \label{fig:Venn}
\end{figure}

\begin{figure*}
    \centering
    \includegraphics[width=\textwidth]{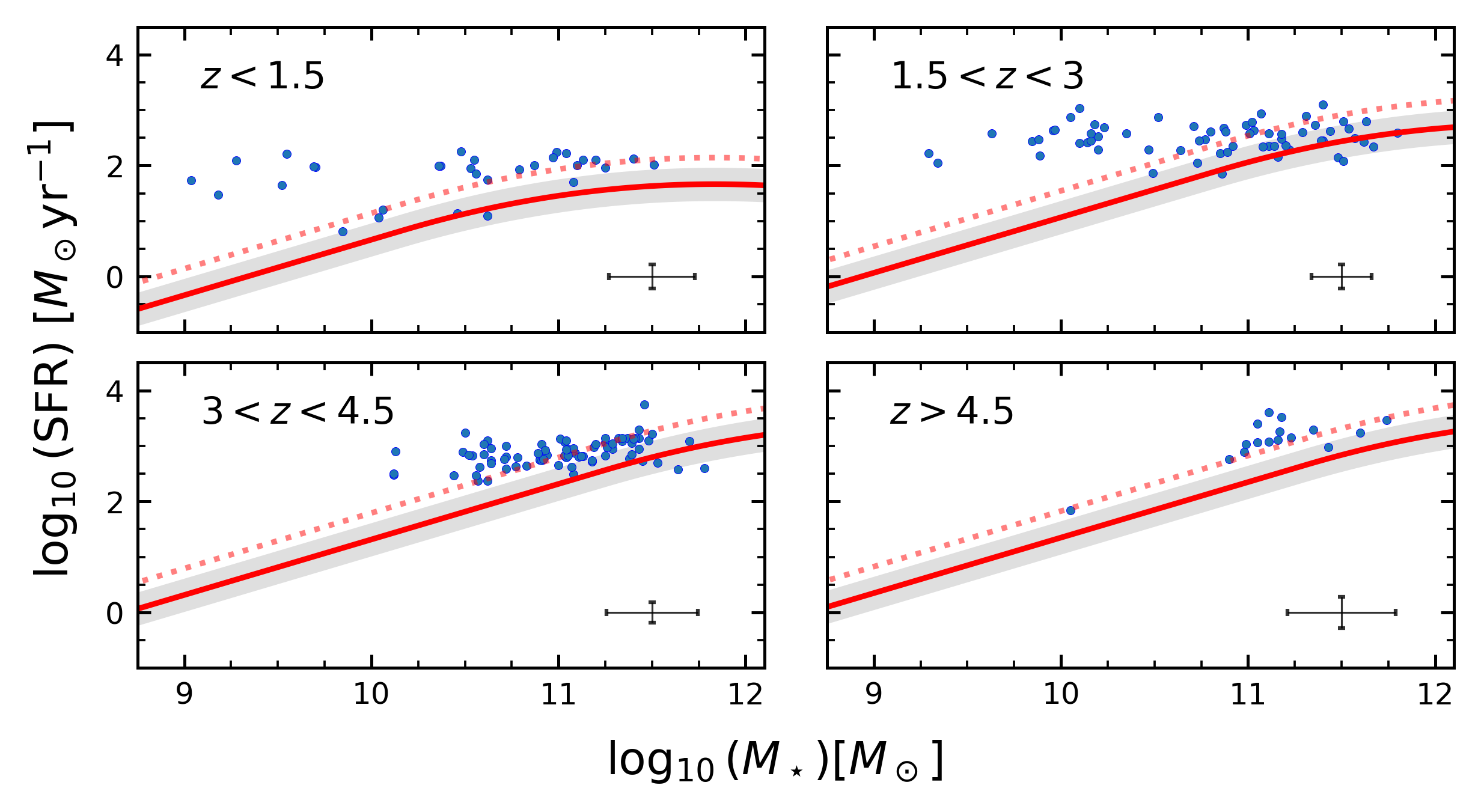}
    \caption{Comparison between the SFRs and stellar masses of the RS-NIRdark galaxies (as computed by \textsc{Magphys}) and those expected from the main sequence of the star-forming galaxies (red solid line; \citealt{Schreiber_15} rescaled to a \citealt{Chabrier_03} IMF). The gray shaded area represents the $1\sigma=0.3$ { dex scatter, while the red dotted line reports our threshold for selecting star-burst galaxies (i.e. three times the SFR expected from MS galaxies)}. It is possible to notice how the vast majority of the galaxies in the sample lie above the main-sequence line. { The median uncertainty on the SFR and on the stellar mass is reported in the lower corner of each panel}. Further details are given in \Sec\ref{sec:analysis_prop}.}
    \label{fig:MS}
\end{figure*}

\section{Discussion}
\label{sec:discussion}
\subsection{Analysis of the physical properties}
\label{sec:analysis_prop}

The results shown in \Fig\ref{fig:Properties} and \Tab\ref{tab:properties}, allow us to infer some general properties of the RS-NIRdark galaxies: 
\begin{itemize}

\item The high value of the median stellar attenuation $A_V\sim3.5$ confirms the initial hypothesis that our selection can provide a sample of highly-obscured galaxies.
\item The high value of { the infrared luminosity ($L_{IR}>10^{12}L_\odot$) and - hence -} the high median SFR ($\sim300$ M$_\odot$yr$^{-1}$) suggests that our galaxies are actively forming stars (as expected from a radio selection). To determine the nature of the RS-NIRdark galaxies, we need to compare the SFRs obtained through the SED-fitting with those expected from main sequence galaxies in the same redshift range \citep[e.g.][]{Schreiber_15}. As shown in \Fig\ref{fig:MS}, most galaxies lie above the main sequence line at all the redshifts. Moreover, when computing the $SFR/SFR_{MS}$ (i.e. the ratio between the SFR and that expected from a main sequence galaxy of the same mass and in the same redshift bin), we obtain that more than 50\% of the RS-NIRdark galaxies have a $SFR/SFR_{MS}>3$, being candidate starburst galaxies.
\item An additional result, strictly connected to the previous one, concerns the high median stellar mass estimated through SED-fitting (M$_\star\sim10^{11}{\rm M}_\odot$). This quantity - albeit quite uncertain due to the weak constraints in the optical/NIR wavelengths - suggest that the RS-NIRdark galaxies are a population of massive star-forming galaxies, { with a high-\textit{z} tail } suitable for playing a significant role in the evolution of the massive and passive galaxies at $z\sim3$ \citep[see e.g.][]{Straatman_14,Schreiber_18,Valentino_20}

\end{itemize}

All these findings, once combined, confirm the initial hypothesis that the RS-NIRdark galaxies represent a significant population of high-\textit{z} DSFGs.

\subsection{ High-z tail}
\label{sec:highz}

The redshift distribution of the RS-NIRdark galaxies is worth a deeper discussion. The median redshift around $z\sim3$ tells us that we are looking at a population whose bulk is located at the so-called \textit{cosmic noon}. However, the presence of a significant tail of high-\textit{z} sources (namely, 99 galaxies at $z>3$ and 17 galaxies at $z>4.5$, { once excluded the possible AGNs selected in \Sec\ref{sec:AGN}}.) can provide some insights on the possible evolutionary path of these sources. 

The { main} result concerns the number density of these sources. We compute this quantity through the $V_{\rm max}$ method \citep{Schmidt_68}, considering the galaxies { located at $z>3.5$ and that could - therefore - play a role in the evolution of the massive galaxies at $z\sim3.5$. We account for the uncertainties in the photo-\textit{z}s and in the radio fluxes through a MonteCarlo integration. Specifically, we perform a large number of realizations ($\sim 500$) of the total number density, sampling each time and for each source a couple of values for the redshift and for the radio flux from their distribution (namely, the $p(z)$ computed by \textsc{Magphys} and the values and uncertainties reported in \citealt{Smolcic_17}). At the end of this procedure, we consider as our number density the median value of the distribution and as its uncertainty the symmetrized interval between the 16th and the 84th percentiles. We obtain a number density of $n=(3.3\pm0.9)\times10^{-6}$ ${\rm Mpc}^{-3}$ for the galaxies at $z>3.5$. This quantity is by a factor 6 lower than those computed by \citet{Straatman_14}, \citet{Schreiber_18}, and \citet{Valentino_20} for the passive and massive galaxies at $z\sim3.5$ ($\sim 2\times10^{-5}$ Mpc$^{-3}$). It is important to notice - however - that the our estimation must be considered as a lower limit on the actual number density of the RS-NIRdark galaxies, since it is not corrected for the expected \textit{duty cycle} of the galaxies and for the incompleteness of the selection. This issue will be discussed in detail in the forthcoming paper of the series (Gentile et al., in prep).} { Moreover, when looking at the stellar masses of our sources, we obtain that the RS-NIRdark galaxies located at $z>3.5$ have a median stellar mass of $\log(M_\star)=11.0 M_\odot$ with a $1\sigma$ dispersion of $0.45$ dex. Comparing this quantity with the expected properties of the progenitors of the massive galaxies at $z\sim3.5$ (see e.g. the forecasts by \citealt{Valentino_20}), we can notice how the low-mass end of those objects cannot be formed by the galaxies in our sample. This result also can be used to explain the difference between the number densities of the two populations.}


\begin{figure}
    \centering
    \includegraphics[width=\columnwidth]{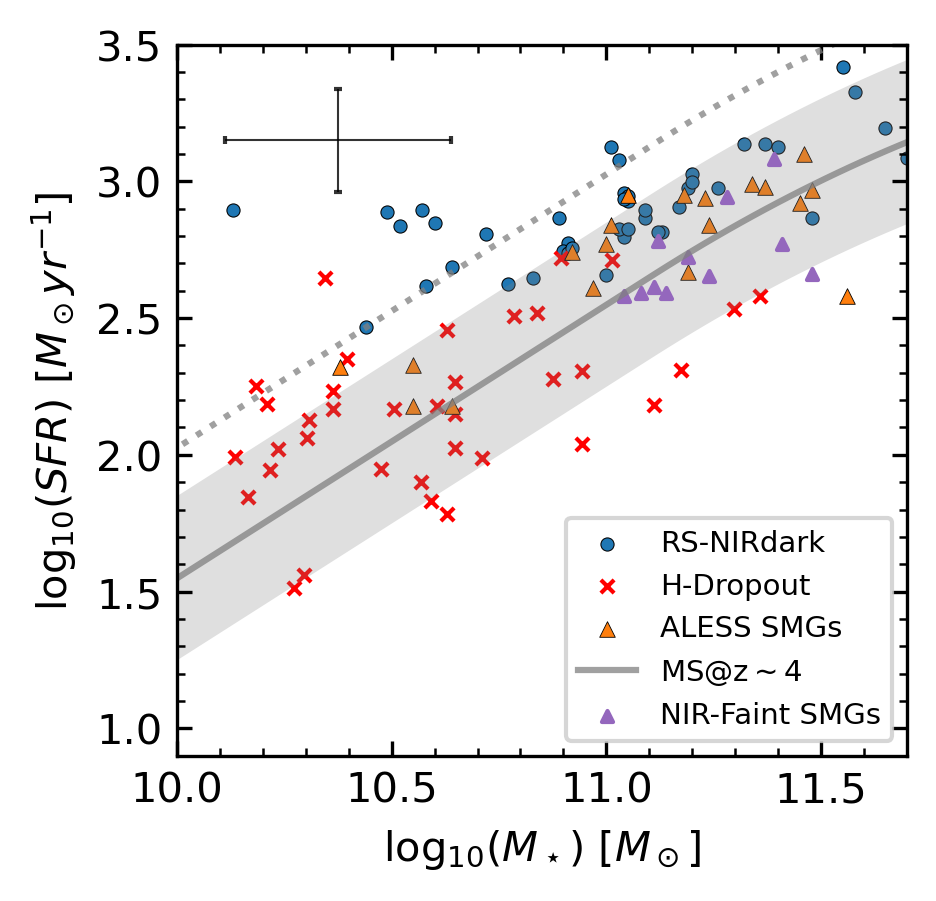}
    \caption{Comparison between the SFRs and stellar masses of several populations of DSFGs. The different symbols show the RS-NIRdark galaxies (blue dots), the H-dropout (\citealt{Wang_19}; red crosses), the SMGs found in ALESS (\citealt{daCunha_15}; orange triangles) and the NIR-faint SMGs discovered by \citet{Smail_21} in the UKIDSS Ultra Deep Survey (purple triangles). The solid gray line represents the main sequence at $z\sim4$ (\citealt{Schreiber_15} rescaled to a \citealt{Chabrier_03} IMF) and the gray shaded area its $1\sigma=0.3$ dex { scatter}. The dotted gray line reports our threshold to define starburst galaxies (i.e. three times the SFR of a main sequence gaalxy). { The median uncertainty on the SFR and on the stellar mass is reported in the lower corner of each panel. For consistency with the cited studies, we only report galaxies with $3.5<z<4.5$.} Further details are given in \Sec\ref{sec:analysis_prop}.}
    \label{fig:comparison_MS}
\end{figure}

\subsection{Comparison with the literature}
\label{sec:comparison_other}

An interesting final point to discuss concerns the RS-NIRdark galaxies and their possible overlap with other important populations of DSFGs: 
\begin{itemize}
    \item \textbf{H-dropout:} The first population is composed of the H-dropout galaxies \citep{Wang_19}. These galaxies are selected as H-dark sources in the CANDELS survey (the limiting magnitude at $5\sigma$ in the H-band is 27mag; \citealt{Grogin_11,Koekemoer_11}) with a counterpart in the second channel of the IRAC camera from the SEDS survey ([4.5]$<$24mag; 80\% completeness limit; \citealt{Ashby_13}). When considering the full photometric catalog of 323 RS-NIRdark galaxies, we obtain that only 266 sources satisfy the selection criteria exposed in \citet{Wang_19}, { the others being too faint at 4.5 $\mu m$ to be { selected with the cut on the [4.5] flux performed by \citet{Wang_19}}}. This result indicates that a significant fraction of the RS-NIRdark galaxies would not be selected as H-dropout. On the contrary, when examining the 18 H-dropout galaxies selected by \citet{Wang_19} in the COSMOS field, we find that only 2 sources have a significant ($S/N>5.5$) radio counterpart at 3 GHz in the catalog by \citet{Smolcic_17}. These two results, when combined, ensures that the two selections of RS-NIRdark galaxies and H-dropout are different, with just some sources belonging to both.
    { \item \textbf{SMGs:} The second population is composed of the so-called sub-millimeter galaxies (SMGs). Since this definition can be applied to all the sources detected in a (sub)mm survey, it strongly depends on the considered instrument's sensitivity. To obtain results comparable with those reported by \citet{Wang_19}, we consider the ALESS survey \citep{Swinbank_14} targeting bright galaxies in the SCUBA survey (i.e. galaxies with a flux density at $870\mu$m $S_{870}>4.2$ mJy). We can estimate the (sub)mm flux density of the RS-NIRdark galaxies at $870\mu$m through the best-fitting SED provided by \textsc{Magphys}. In doing so, we obtain that only 76 galaxies have $S_{870}>4.2$ { mJy}, being consistent with the selection by \citet{Swinbank_14}. On the other side, \citet{Thomson_14} pointed out that only $\sim$70\% of SMGs are radio-bright galaxies, while \citet{Gruppioni_20} pointed out that only a tiny percentage of SMGs are NIR-dark galaxies. As before, these results suggest that the RS-NIRdark galaxies and SMGs are two different populations of galaxies with just some sources in common.}
\end{itemize}

{ Finally, we can compare the physical properties estimated by \citet{Wang_19} and \citet{daCunha_15} for the population of H-dropout and SMGs with those presented in this study. \Fig\ref{fig:comparison_MS} shows the comparison between the SFR and stellar mass of the RS-NIRdark galaxies in the redshift range $3.5<z<4.5$ (i.e. around the median redshift $z\sim4$ reported by \citealt{Wang_19} for the H-dropout) with those reported by \citet{daCunha_15} and \citet{Wang_19} and with the main sequence at this redshift (\citealt{Schreiber_18b} rescaled to a \citealt{Chabrier_03} IMF). For completeness, we also include in the plot the sample of “NIR-faint" ($Ks>23.9$) SMGs selected in COSMOS by \citet{Smail_21}.

We can notice that the RS-NIRdark galaxies are - on average - { more star-forming than the H-Dropout ($\Delta \log$(SFR/M$_\odot$ yr$^{-1}$)$\sim$0.6), with a median SFR comparable with the SMGs ($\Delta \log$(SFR/M$_\odot$ yr$^{-1}$)$\sim$0.02),} but with a significant tail of less star-forming sources. This result can be explained by the radio-selection, cutting out most of the sources below the main sequence (belonging { mainly} to the H-dropout). Additionally, we can see how the RS-NIRdark galaxies cover most of the mass range covered by the other selections, { being on average more massive than the H-dropout and less massive than the SMGs}. This result is in agreement with the discussed overlap with both the other populations.}

\section{Summary}
\label{sec:summary}

In this paper, we presented the first panchromatic study of { 263} Radio-Selected NIRdark galaxies discovered in the COSMOS field following the selection by \citet{Talia_21}. The development of a new deblending tool (\textsc{PhoEBO}: Photometry Extractor for Blended Objects) allowed us to extract accurate photometry in the optical-to-MIR regime, even for the sources with a close bright contaminant. This procedure, in particular, allowed us to analyze a wider sample of galaxies missed in previous studies on the RS-NIRdark galaxies in the COSMOS field \citep{Talia_21,Behiri_23}. The complete photometric catalog\footnote{The complete catalog, together with the other materials supporting the findings of this study are available on the website of the collaboration: \url{https://sites.google.com/inaf.it/rsnirdark/}} has been employed to estimate the photo-\textit{z}s and physical properties of all the galaxies in the sample through an SED-fitting procedure performed with two complementary codes (\textsc{Magphys} and \textsc{Cigale}). The results obtained with these algorithms confirmed the initial hypothesis that the RS-NIRdark galaxies are a population of starburst DSFGs, lying above the main sequence in all the redshift bins and with a significant amount of dust absorbing their optical/NIR emission. Moreover, by studying in detail the redshift distribution of the galaxies in the sample and their number density, we obtain precious insights on the possible evolutionary path of these sources, collecting significant clues that the RS-NIRdark galaxies could play a key role in the evolution of the massive and passive galaxies discovered at $z\sim3$. In addition, the analysis of the multi-wavelength counterpart in all the wavelength regimes (from the X-rays to the radio) allowed us to estimate the possible AGN contribution in our sample. Finally, through a comparison with other populations of DSFGs, we confirmed that the radio-selection produces a population of galaxies with different physical properties with respect to the SMGs \citep[e.g.][]{daCunha_15} and the H-Dropout \citep{Wang_19}.

The catalog presented in this work will be employed in the next papers of the series to derive the luminosity function of the RS-NIRdark galaxies, to estimate their contribution to the cosmic SFRD and to study their role in the evolution of the massive galaxies. In the near future, the results presented in this work will be reinforced by the plethora of new data coming from state-of-the-art facilities. About half of the RS-NIRdark galaxies will be observed by the James Webb Space Telescope as a part of the COSMOS-Web survey \citep{Casey_22}. Moreover, a first spectral analysis of a pilot sample of these galaxies performed with ALMA will allow a robust determination of the spectroscopic redshifts (Gentile et al, subm.).

\begin{acknowledgments}
{ We warmly thank the anonymous referee for his/her comments on the first version of this paper, which really allowed us to increase the overall quality of our study.}
FG and MT thank Eleni Vardoulaki for sharing her sample of AGN in COSMOS. FG and MT thank Ian Smail for sharing his sample of NIR-faint SMGs in COSMOS. FG thanks Alberto Traina for the valuable support on the SED-fitting. FG, MT, AL, MM and AC acknowledge the support from grant PRIN MIUR 2017-20173ML3WW\_001. ‘Opening the ALMA window on the cosmic evolution of gas, stars, and supermassive black holes’. \\
AL is supported by the EU H2020-MSCA-ITN-2019 project 860744 ‘BiD4BEST: Big Data applications for Black hole Evolution STudies’. MV acknowledges financial support from the Inter-University Institute for Data Intensive Astronomy (IDIA), a partnership of the University of Cape Town, the University of Pretoria, the University of the Western Cape and the South African Radio Astronomy Observatory, and from the South African Department of Science and Innovation's National Research Foundation under the ISARP RADIOSKY2020 Joint Research Scheme (DSI-NRF Grant Number 113121) and the CSUR HIPPO Project (DSI-NRF Grant Number 121291).
\end{acknowledgments}

\vspace{5mm}

\appendix

\section{Comparison with Talia+21 and Behiri+23}
\label{sec:comparison_t21}

\begin{deluxetable}{ccccc}
\label{tab:properties_otherstudies}
\tablecaption{Comparison between the median properties obtained through \textsc{Magphys} in this paper and in the previous works based on sub-samples of the RS-NIRdark galaxies \citep{Talia_21,Behiri_23}}
\tablewidth{0pt}
\tablehead{
\colhead{Property} & \colhead{This Work} & \colhead{Talia+21} & \colhead{Behiri+23} & \colhead{Unit}
}
\startdata
$z_{\rm phot}$ & $2.96\pm 0.04$ & $3.1\pm0.1$& $3.3\pm0.2$ &\\
$\log(M_\star)$ &$11.01 \pm 0.02$ & $11.3\pm0.02$ & $11.1\pm0.02$ &$M_\odot$ \\
$\log(SFR)$ &$2.67 \pm 0.02$ & $2.48\pm0.02$ & $2.79\pm0.02$ &$M_\odot {\rm yr}^{-1}$ \\
$A_{\rm v}$ & $4.01\pm0.04$& $4.2\pm0.1$ & $4.0\pm0.03$& mag\\
$\log(L_{\rm Dust})$ & $12.45\pm0.02$ & $12.50\pm0.02$ & $12.93\pm0.02$ & $L_\odot$ \\
$T^L_{\rm Dust}$ &  $42.0\pm0.2$& $39.5\pm0.7$ & - & K\\
$\log(M_{\rm Dust})$&$8.57 \pm 0.02$& -- & -- &$M_\odot$
\enddata 
\end{deluxetable}

As previously discussed in \Sec\ref{sec:intro}, this paper continues the previous studies on the RS-NIRdark galaxies in the COSMOS field started by \citet{Talia_21} and \citet{Behiri_23}. Here, we discuss the main differences between this work and the previous ones. The main upgrade of this work concerns the selection of the sample of RS-NIRdark galaxies. Firstly, by cross-matching the radio catalog with the NIR-selected COSMOS2020 (in place of the COSMOS2015 employed by \citealt{Talia_21} and \citealt{Behiri_23}), we have been able to exclude from the sample 153 NIR-faint galaxies missed in the previous version of the COSMOS catalog. These galaxies were undetected in the COSMOS2015 because of the brighter limiting magnitude reached in the NIR bands ($\sim 1$ mag in the $Ks$ band) and in the shallower $\chi^2$-image employed as the detection image \citep[see][]{Laigle_16,Weaver_22}. Secondly, both \citet{Talia_21} and \citet{Behiri_23} employed an additional selection step after the three discussed in this paper (\Sec\ref{sec:selection}). These step were aimed to only analyze the sources without contamination from a nearby source. More in detail, \citet{Talia_21} removed from the sample all the galaxies with a nearby source whose $3\sigma$ isophote in the $\chi^2$-image overlapped the $3\sigma$ isophote in the radio map at 3 GHz. Thanks to this selection, \citet{Talia_21} analyzed a sub-sample of the full catalog composed by 197 galaxies. Similarly, \citet{Behiri_23} slightly enlarged the sample by including 75 “slightly contaminated" sources (i.e. galaxy with a NIR-bright contaminant overlapping the $5\sigma$ radio isophote). Differently from these works, in this paper we developed an \textit{ad-hoc} pipeline (\textsc{PhoEBO}; see \Sec\ref{sec:photometry_extraction}) to extract the photometry of all the galaxies, regardless of the possible blending with nearby contaminants.
Other important differences are:
\begin{itemize}
    \item Differently from \citet{Talia_21} and \citet{Behiri_23}, in this paper we employ the IRAC maps from the DAWN survey (\Sec\ref{sec:magphys}). These maps are significantly deeper than those (from the SPLASH survey\footnote{\url{https://splash.caltech.edu/index.html}}) used in the previous studies, allowing us to obtain more detection in the IRAC bands for our galaxies and put more stringent constraints on the upper limits for the IRAC-undetected sources. Similarly, the flux densities in the (sub)mm regime are obtained from the most up-to-date version of the A3COSMOS catalog, reflecting in a more significant number of counterparts for our sample of galaxies. 
    
    \item Finally, In \citet{Talia_21}, the photometry is extracted \textit{ex-novo} only for the IRAC bands, while the flux densities and the upper limits in the optical and NIR bands are obtained from pre-existing catalogs or the average depth of the analyzed maps. In this work, the pipeline presented in \Sec\ref{sec:photometry_extraction} allows us to perform forced photometry on the exact position of the sources (thanks to the radio prior). This procedure (analogous to that followed by \citealt{Behiri_23} for the NIR-dark galaxies) enable us to pose more stringent upper limits in the optical and NIR band, reducing the degrees of freedom in the SED-fitting. 
\end{itemize}

These improvements allow us to build a more complete photometric catalog employed in the SED-fitting procedure. Analyzing the results obtained with \textsc{Magphys} (i.e. the same software used in the previous works; see \Tab\ref{tab:properties_otherstudies}), we can notice some interesting differences: 
\begin{itemize}
    \item The median redshift of the whole sample of RS-NIRdark galaxies is slightly lower (see \Tab\ref{tab:properties_otherstudies} than that reported in \citet{Talia_21} and \citet{Behiri_23}.
    \item The median stellar attenuation is slightly lower than that reported in the previous studies (\Tab\ref{tab:properties_otherstudies}). This result can be explained by the more stringent upper limits obtained by the new-deblending algorithm in the optical and NIR bands and in the MIR regime thanks to the deeper IRAC maps employed in this work. 
\end{itemize}

 { The main difference between the results of the various studies concerns the number density of the RS-NIRdark galaxies. Following the procedure discussed in \Sec\ref{sec:highz}, we can compute this quantity for the galaxies located at $z>4.5$ (i.e. the highest redshift bin adopted in \citealt{Talia_21}). We obtain $n=(1.1\pm0.6)\times10^{-6}$ ${\rm Mpc}^{-3}$ . This quantity is in moderate tension with that obtained by \citet{Talia_21}: $n=(5.2\pm1.3)\times10^{-5}$ ${\rm Mpc}^{-3}$. This result can be explained by the lower number of galaxies analysed in this study with respect to the previous one and with the lower median redshift obtained here.}.

 \section{Data Release}

 { The full photometric catalog of the RS-NIRdark galaxies in the COSMOS field, together with the other data supporting this study can be freely accessed on the website of the collaboration\footnote{\url{https://sites.google.com/inaf.it/rsnirdark/}}. More in detail, the data release consists in:
 \begin{itemize}
     \item \textbf{One catalog containing all the photometry from optical to radio wavelengths:} The catalog has follows this structure:
     \begin{itemize}
         \item \textbf{Column 1}: A progressive ID. For consistency with the previous studies by \citet{Talia_21} and \citet{Behiri_23}, the IDs refer to the total sample of 476 RS-NIRdark galaxies selected in the COSMOS field by \citet{Talia_21}.
         \item \textbf{Columns 2-27}: Fluxes and photometric uncertainties obtained through the \textsc{PhoEBO} for all the bands discussed in \Tab\ref{tab:maps}.
         \item \textbf{Columns 28-85}: Fluxes and photometric uncertainties at FIR/(sub)mm/radio wavelengths obtained through cross-matching with the pre-existing catalogs listed in \Sec\ref{sec:additional}.
     \end{itemize}
     All the fluxes and uncertainties are reported in Jy. The catalog can be directly employed to perform the SED-fitting with \textsc{Magphys}, therefore all the entries with a $S/N<1$ are treated as upper limits (i.e. with a missing flux and with a positive uncertainty equal to the $1\sigma$ upper limit).
     \item \textbf{Two catalogs containing all the properties estimated through SED-fitting with \textsc{Magphys} and \textsc{Cigale}:} A complete description of all the columns can be found in in the FITS header of each file. 
    \item \textbf{A catalog with the AGN classifications performed in \Sec\ref{sec:AGN}:} The description of the columns is contained in the FITS header of the file.
     \item \textbf{A series of pdf files showing the best-fitting SEDs of all the RS-NIRdark galaxies as computed by \textsc{Magphys} and \textsc{Cigale} and the cutouts in most of the bands analysed with \textsc{PhoEBO}}
 \end{itemize}}

\bibliography{sample631}{}
\bibliographystyle{aasjournal}

\end{document}